\documentclass[fleqn,usenatbib]{mnras}

\usepackage{newtxtext,newtxmath}

\usepackage[T1]{fontenc}

\DeclareRobustCommand{\VAN}[3]{#2}
\let\VANthebibliography\thebibliography
\def\thebibliography{\DeclareRobustCommand{\VAN}[3]{##3}\VANthebibliography}

\usepackage{graphicx}	
\usepackage{amsmath}	

\usepackage{caption}
\usepackage{subcaption}

\usepackage{cuted}

\DeclareMathOperator{\sgn}{sgn}

\title[Focusing of Nonlinear Eccentric Waves. II.]{Focusing of nonlinear eccentric waves in astrophysical discs. II. Excitation and damping of tightly-wound waves.} 

\author[E. M. Lynch]{
Elliot M. Lynch$^{1,2}$ \thanks{E-mail: elliot.lynch@ens-lyon.fr}
\\
$^{1}$Department of Applied Mathematics and Theoretical Physics, University of Cambridge, Centre for Mathematical Sciences, \\ Wilberforce Road, Cambridge CB3 0WA, UK\\
$^{2}$Univ Lyon, Univ Lyon1, Ens de Lyon, CNRS, Centre de Recherche Astrophysique de Lyon UMR5574, F-69230, Saint-Genis,-Laval, France.\\
}

\date{Accepted XXX. Received YYY; in original form ZZZ}

\pubyear{2021}

\begin{document}
\label{firstpage}
\pagerange{\pageref{firstpage}--\pageref{lastpage}}
\maketitle

\begin{abstract}
In this paper I develop a nonlinear theory of tightly-wound (highly twisted) eccentric waves in astrophysical discs, based on the averaged Lagrangian method of Whitham. Viscous dissipation is included in the theory by use of a pseudo-Lagrangian. This work is an extension of the theory developed by Lee \& Goodman to 3D discs, with the addition of viscosity.  I confirm that linear tightly-wound eccentric waves are overstable and are excited by the presence of a shear viscosity and show this persists for weakly nonlinear waves. I find the waves are damped by shear viscosity when the wave become sufficiently nonlinear, a result previously found in particulate discs. Additionally I compare the results of this model  to recent simulations of eccentric waves propagating in the inner regions of black hole discs and show that an ingoing eccentric wave can be strongly damped near the marginally stable orbit, resulting in a nearly circular disc with a strong azimuthal variation in the disc density.
\end{abstract}




\section{Introduction}

Astrophysical discs can be eccentric in general, with the dominant motion of the fluid consisting of a sequence of confocal elliptical Keplerian orbits. In a fluid disc these elliptical orbits evolve slowly due to the action of external perturbations, pressure gradients, self-gravity and turbulent stresses. When the eccentricity evolves slowly on the disc precession timescale, it is expected to be well approximated by an eccentric mode. An eccentric mode is an untwisted eccentric disc which is stationary in a frame that rotates at the constant disc precession frequency; these represent the eccentric standing waves of the disc. The eccentric modes in non-self-gravitating discs have been extensively studied in both the linear \citep{Kato83,Okazaki97,Okazaki91,Papaloizou02,Papaloizou06,Goodchild06,Ogilvie08,Teyssandier19} and nonlinear \citep{Barker16,Ogilvie19,Lynch19,Zanazzi20} regimes, as they are the simplest solutions to ideal eccentric disc theory. 

The modes considered in \citet{Lynch19} (henceforth Paper I) were untwisted. These are the superposition of inward and outward travelling eccentric waves which reflect from the boundaries and carry equal and opposite angular momentum fluxes and twists. When these are superposed the angular momentum flux and twists cancel, producing the untwisted standing wave. If there is some difference in the amplitudes of the ingoing and outgoing waves then the angular momentum flux (and thus twist) doesn't perfectly cancel resulting in a twisted disc. This can happen if one of the boundaries isn't perfectly reflecting allowing one of the waves to be partially or completely transmitted, as seen in the simulations of \citet{Dewberry19b} and \citet{Dewberry20}. Viscosity can also cause discs to become twisted by damping the reflected wave so that it is of lower amplitude than the incoming wave; this was explored by \citet{Ferreira09} in the context of the inner region of black hole discs. In this case the eccentric wave is typically untwisted on the boundary as the incoming and outgoing waves have the same amplitude there, but develops a significant twist far from it \citep{Goodchild06,Ferreira09,Teyssandier16,Teyssandier17}. 

In most eccentric discs the eccentricity is expected to be excited at a resonance with an orbital companion \citep{Lubow91} and propagates away to be damped elsewhere in the disc. \citet{Goodchild06} studied the effect of resonances and bulk viscosity in 2D linear theory. \citet{Ferreira09} considered the influence of bulk viscosity on an eccentric wave reflecting off a free boundary. The interaction of discs and planets in a viscous disc was studied by \citet{Teyssandier16,Teyssandier17} which results in a moderately twisted eccentric wave in the disc generated by the planets. Simulations of this effect were carried out by \citet{Rosotti17} and \citet{Ragusa18}. \citet{Dewberry19b} and \citet{Dewberry20} studied eccentric waves in a pseudo-Newtonian disc that were excited at the disc boundaries and propagate towards the marginally stable orbit as a highly twisted eccentric wave. In general the twisting of eccentric waves is important whenever eccentric waves are generated or damped within the discs, or when they are free to leave the domain. 

The inclusion of dissipation in eccentric disc theory is complicated by the fact that, under certain circumstances, eccentricity can be excited by the presence of a shear viscosity \citep{Syer92,Syer93,Lyubarskij94,Ogilvie01,Latter06}. Notably a circular disc is unstable to the excitation of eccentric waves in the presence of an shear $\alpha$-viscosity, commonly used as a model of disc turbulence. This excitation is a form of viscous overstability, originally discovered by \citet{Kato78} for axisymmetric perturbations and previously studied in the context of planetary rings \citep{Goldreich78b,Borderies85,Borderies86}. 

One can model dissipation of eccentric waves using a bulk viscosity (e.g. in \citet{Ferreira09}), as this always damps eccentricity. Alternatively it has been suggested in \citet{Ogilvie01} that the finite response time of the turbulence can suppress the overstability. Of course one can sidestep this issue entirely by explicitly modelling disc turbulence, typically sourced from the magnetorotational-instability (MRI). However, the picture has been further complicated by the recent discovery that a sufficiently nonlinear eccentric wave is capable of suppressing the MRI \citep{Dewberry20}.

In this paper we apply the theory of \citet{Whitham65} and \citet{Jimenez76} to determine the dynamics of these tightly-wound eccentric waves. These waves occur when a travelling eccentric wave encounters strong apsidal precession, as might occur in the inner regions of black hole discs. Our work is an extension of theory of tightly-wound eccentric waves developed by \citet{Lee99} to include viscosity and a more general treatment of pressure forces that takes into account the dynamical vertical structure of the disc. Tightly-wound waves are highly twisted, short-wavelength, eccentric waves; untwisted short-wavelength waves were studied in Paper I. In principle the untwisted short-wavelength theory developed in Paper I and the tightly-wound theory developed here could be obtained as limiting cases of a more general twisted-short wavelength theory, but we shall not attempt that here.

A version of the work presented here appeared in the authors PhD thesis \citet{LynchPhd}. The present work incorporates several novel effects that are important for describing self-gravity and viscosity in 3D eccentric discs. This work also improves on the comparison with \citet{Dewberry19b,Dewberry20}.

This paper is structured as follows. In Section \ref{eccentric disc prelim} we discuss the geometry of eccentric discs and introduce the Hamiltonian formalism derived in \citet{Ogilvie19}. In Section \ref{2.5D overview} we derive the Lagrangian for a tightly wound eccentric wave. In Section \ref{conservative dynamics gen and damp} we consider the conservative dynamics of tightly wound waves in the absence of dissipation. Section \ref{modulation} confirms that the wave-steepening conditions derived in Paper I for untwisted, short wavelength, waves apply to tightly wound waves. Section \ref{pseudo lagrangian prescription} shows how viscosity can be included by use of a pseudo-Lagrangian prescription, with Section \ref{nonconservative wave dynamics} presenting the dynamics of a tightly wound eccentric wave subject to precessional forces, pressure gradients, self gravity, bulk and shear viscosity. In Section \ref{pseudo-Newtonian disc discussion} we discuss an application of our theory to the pseudo-Newtonian discs considered by \citet{Dewberry19b} and \citet{Dewberry20}. Conclusions are given in Section \ref{conc} and mathematical derivations are given in the appendices.

\section{Nonlinear Theory of Eccentric Discs} \label{eccentric disc prelim}

\subsection{Eccentric orbital coordinate system}

Let $(r,\phi)$ be polar coordinates in the disc plane. The polar equation for an elliptical Keplerian orbit of semimajor axis $a$, eccentricity $e$ and longitude of periapsis $\varpi$ is

\begin{equation}
r = \frac{a (1 - e^2)}{1 + e \cos f} \quad ,
\end{equation} 
where $f = \phi - \varpi$ is the true anomaly. Equivalently,

\begin{equation}
r = a (1 - e \cos E) ,
\end{equation} 
where $E$ is the eccentric anomaly which satisfies

\begin{equation}
\cos f = \frac{\cos E - e}{1 - e \cos E}, \quad \sin f = \frac{\sqrt{1 - e^2} \sin E}{1 - e \cos E}\quad .
\end{equation}

A planar eccentric disc involves a continuous set of nested elliptical orbits. The shape of the disc can be described by considering $e$ and $\varpi$ to be functions of $a$. The derivatives of these functions are written as $e_a$ and $\varpi_a$, which can be thought of as the eccentricity gradient and the twist, respectively. The disc evolution is then described by the slow variation in time of the orbital elements $e$ and $\varpi$ due to secular forces such as pressure gradients in the disc and secular gravitational interactions.

We now introduce the $(a,E)$ orbital coordinate system of \citet{Ogilvie19}. In this coordinate system $a$ is an orbit labelling coordinate that specifies an orbit, while $E$ denotes where on the orbit the point is located. The shape of the orbit is determined by $e$ and $\varpi$ which are functions of $a$ and time. The Jacobian determinant of this coordinate system can be written in the form $J = a j (1 - e \cos E)$, where

\begin{equation}
j = \frac{1 - e (e + a e_a)}{\sqrt{1 - e^2}} (1 - q \cos(E -  E_0)) ,
\end{equation}
and we have introduced the nonlinearity parameter $q$, defined by

\begin{equation}
q^2 = \frac{(a e_a)^2 + (1 - e^2) (a e \varpi_a)^2}{[1 - e (e + a e_a)]^2} \quad ,
\label{nonlinearity def}
\end{equation}
We require $|q| < 1$ to avoid an orbital intersection. The relative contribution of the eccentricity gradient and twist to $q$ is determined by an angle $E_0$ with 

\begin{equation}
\frac{a e_a}{1 - e(e + a e_a)} = q \cos E_0 \, ,\quad \frac{\sqrt{1 - e^2} a e \varpi_a}{1 - e(e + a e_a)} = q \sin E_0 \, . 
\end{equation}

\subsection{Eccentric disc Lagrangian}

\citet{Ogilvie19} derived the Hamiltonian density for a 3D fluid disc, where the only non-Keplerian terms are from pressure gradients,

\begin{equation}
H_{a} = \frac{1}{2} (\gamma + 1) M_a \langle \bar{\varepsilon} \rangle \quad ,
\end{equation}
where $M_a = dm/da$ is the one-dimensional mass density with respect to $a$ and $\bar{\varepsilon}$ is the mass-weighted vertically averaged specific internal energy. Angle brackets denote an orbit average. Both the 2D and 3D Hamiltonian densities can be written as $H_a = H^{\circ}_a F$ where $H^{\circ}_a = 2 \pi a P^{\circ}$ is the Hamiltonian of a circular disc, with $P^{\circ}$ the radial profile of the vertically integrated pressure in the limit of a circular disc, and $F = F(e, a e_a, a \varpi_a)$ is a dimensionless function that depends on the disc geometry \citep{Ogilvie19}.

Additional precessional forces can be included by adding a term to the Hamiltonian that depends on $a$ and $e^2$ only: $H^{f}_{a} = H^{f}_a (a,e^2)$, so that the Hamiltonian density is given by $H_{a}  + H^{f}_{a}$.
 
There are two sources of nonlinearity present in the Hamiltonian density. One is associated with the eccentricity $e$; the other is associated with the gradients of the solution and typically determines how oscillatory (or twisted) the wave is, with more oscillatory (or twisted) waves being more nonlinear. This nonlinearity is characterised by the nonlinearity or orbital intersection parameter $q$ given by Equation \ref{nonlinearity def}. We can write the  geometric part of the Hamiltonian in terms of $e$, $q$ and $E_0$ so that $F = F(e,q,E_0)$.

From the Hamiltonian for the disc dynamics we can obtain the associated Lagrangian via a non-canonical Legendre transform,

\begin{equation}
L = \int M_a n a^2 \varpi_t \sqrt{1 - e^2} \, d a - H,
\end{equation}
which leads to a Lagrangian of the form

\begin{equation}
L = \int \left( M_a n a^2 \varpi_t \sqrt{1 - e^2} - H_a - H_a^{f} \right) \, d a ,
\label{full theory Lagrangian}
\end{equation}
where $\varpi_t = \frac{\partial \varpi}{\partial t}$ and $n = (G M_1/a^3)^{1/2}$ is the mean motion with $M_1$ the central mass. Variation of this Lagrangian with respect to $\varpi$ and $e$ leads to the non-canonical Hamilton's equations derived by \citet{Ogilvie19}. These two formalisms are equivalent; however it will be easier to extend the Lagrangian formalism to include dissipation.

To include disc self gravity we must include an additional contribution $\bar{L}_{\rm sg}$ to the Lagrangian given by Equation \ref{full theory Lagrangian}. An explicit expression for $\bar{L}_{\rm sg}$ in the tight-winding limit is derived in Appendix \ref{self gravity lagrangian derivation}.

\subsection{Dynamical scale height in an eccentric disc}

Unlike circular discs, the vertical structure of an eccentric disc cannot be decoupled from the horizontal fluid motion. Eccentric discs cannot maintain hydrostatic equilibrium owing to the variation of vertical gravity and lateral compression around an elliptical orbit \citep{Ogilvie01,Ogilvie08,Ogilvie14}. The latter of these effects is more important in the short-wavelength limit.

Following \citet{Ogilvie14} we define the disc scale height to be

\begin{equation}
H^2 = \frac{\int \rho z^2 \, d z}{\int \rho \, d z} ,
\end{equation}
where, in this paragraph only, $H$ is the disc scale height. \citet{Ogilvie19} introduced the dimensionless scaleheight $h = H/H^{\circ}$, which is the scale height rescaled by the scale height $H^o(a)$ of a circular disc. In an ideal fluid disc this evolves according to \citep{Ogilvie19}
 
\begin{equation}
\frac{d^2 h}{d t^2} + \frac{n^2 h}{(1 - e \cos E)^3} = \frac{n^2}{j^{\gamma - 1} h^{\gamma}} - \psi_{\rm sg} h ,
\label{scale height equation}
\end{equation}
where $\psi_{\rm sg}$ contains terms due to the disc self-gravity \footnote{Not considered in \citet{Ogilvie19}}. An explicit expression for $\psi_{\rm sg}$, in the tight-winding limit, is given in Appendix \ref{self gravity deriv}.

\section{The time dependent short-wavelength Lagrangian} \label{2.5D overview}

It is convenient to rewrite the Lagrangian \eqref{full theory Lagrangian} as

\begin{align}
\begin{split}
L &= \int M_a n a^2 \tilde{\varpi}_t \sqrt{1 - e^2} \, d a + \bar{L}_{\rm sg} \\
 &+ \int \left[ M_{a} n a^2 \left( \omega \sqrt{1 - e^2} + \int  \frac{e \omega_{f} (a,e) }{\sqrt{1 - e^2}} \, d e \right)  - H^{\circ}_a F \right]\, da ,
\end{split}
\end{align}
which is the Lagrangian for an eccentric wave in a frame rotating at rate $\omega$ and we have introduced the longitude of pericentre in the rotating frame $ \tilde{\varpi} := \varpi - \omega t$. As in  Paper I the precessional forces coming from $H^{f}_a$ can be expressed in terms of the precession rate $\omega_{f} (a,e)$ of a test particle.

We now consider the situation where the precessional (or test particle) forces are strong relative to the background pressure gradient as this gives rise to short-wavelength or tightly wound eccentric waves. Similar to Paper I we define

\begin{equation}
N^2 = \frac{M_a n a^2 ( \omega_{f} (a,0) + \omega_{\rm sg} - \omega)}{H^{\circ}_{a}} \quad ,
\end{equation}
with the limit of interest being $N^2 \gg 1$. Here we have added an additional precessional term,

\begin{equation}
\omega_{\rm sg} = - \frac{1}{2 n a^2} \frac{d}{d a} \left ( a^2 \frac{d \Phi_{l}}{d a} \right) ,
\end{equation}
where $\Phi_{l}$ is the contribution to the disc potential from material at large distances (i.e.\, distances much larger than the disc thickness). This allows us to absorb the long-range contribution to the disc self-gravity into $N^2$. In the tight-winding limit we can approximate $\Phi_{l}$ using the gravitational potential of the reference circular disc, as the asymmetry due to the eccentric wave averages out on long lengthscales. With this definition of $N^2$ the remaining contributions from disc self-gravity are entirely local and can be determined from an eccentric shearing box (See Appendix \ref{self gravity deriv}).

We introduce the rapidly varying phase variable $\varphi = \varphi(a,t)$ which satisfies $a \varphi_a = N(a) k(a)$ with $k(a) = O(1)$ a rescaled dimensionless wavenumber. We also make use of the rescaled variable $\tilde{e} = e N k$; in the limit of interest $e \ll 1$, but $\tilde{e} = O(1)$. The nonlinearity (Equation \ref{nonlinearity def}) becomes

\begin{align}
\begin{split}
q^2 &\approx \left(a \frac{d e}{d a} \right)^2 + a^2 e^2 \left(\frac{d \varpi}{d a} \right)^2 \\
& \approx \tilde{e}^2_{\varphi} (\varphi,a) + \tilde{e}^2 (\varphi,a) \varpi^2_{\varphi} (\varphi,a) \quad ,
\end{split}
\end{align}
which is of order unity. In this limit the Jacobian determinant $J \approx a j$, where $j$ is given by

\begin{equation}
j \approx 1 - q \cos \tilde{E} ,
\end{equation}
and $\tilde{E} = E - E_0$. In the tight-winding limit Equation \ref{scale height equation} simplifies to

\begin{equation}
\frac{d^2 h}{d E^2} + h = \frac{1}{j^{\gamma - 1} h^{\gamma}} - \frac{\psi_{\rm sg} (a,j,h)}{n^2} h ,
\end{equation}
meaning $h$ is an even function of $\tilde{E}$, with $h = h(a,q,\cos \tilde{E})$.

Substituting the expressions for $q$ and $\tilde{e}$ into the Lagrangian and keeping only terms of the lowest order in $N^{-1}$, the Lagrangian becomes

\begin{equation}
L = \int \left[- \frac{M_a n a^2}{N^2} \tilde{\varpi}_t \frac{\tilde{e}^2}{2 k^2}  + H^{\circ}_a  \left(\frac{1}{2} \frac{\tilde{e}^2}{k^2}  - F(q) - \frac{2 \lambda_{\rm sg}}{\bar{Q}^{\circ}} F_{\rm sg} (q) \right) \right] \, da ,
\label{lagrangian in limit}
\end{equation}
where $\bar{Q}^{\circ} = \frac{H^{\circ} n^2}{\pi G \Sigma^{\circ}}$ is the Toomre-Q in the reference disc, $\lambda_{\rm sg}$ is a dimensionless constant of order unity that depends on the disc vertical structure,  and $F_{\rm sg} (q) = \langle j^{-1} h \rangle$ is the geometric part of the local contribution to self-gravity in a 3D disc. This differs substantially from that of a 2D disc (see for instance \citet{Lee99}) because the dominant local contribution to the gravitational self energy is from the suspension of disc material away from the midplane, an effect not present in a 2D disc. A derivation of the local contribution from self-gravity is given in Appendix \ref{self gravity lagrangian derivation}.

We have neglected two leading order terms; $M_{a} n a^2 \omega$ which is a function only of $a$ that has no effect on the dynamics and $M_{a} n a^2 \tilde{\varpi}_t$. It is possible to show that the latter term only contributes when $M_{a} n a^2$ is time dependent, a situation we shall not consider in this paper. A time dependent $M_{a} n a^2$ suggests a redistribution of angular momentum in the disc, which cannot happen in the ideal theory obtained from the Lagrangian \eqref{lagrangian in limit}. We shall later include viscous dissipation, which can redistribute angular momentum, but will neglect this term under the assumption that background disc evolves slowly on the timescales of interest. The neglected $M_{a} n a^2 \tilde{\varpi}_t$ term could be important for a theory that models disc accretion. 

\section{Conservative dynamics of tightly wound waves} \label{conservative dynamics gen and damp}

Initially we neglect non-conservative terms and derive the equations for an ideal, time dependent, tightly wound wave. As the equations are fully conservative this can be done entirely within the framework of Lagrangian mechanics and the averaged Lagrangian theory of \citet{Whitham65}. The short lengthscale dynamics of a tightly wound wave are trivial, with $\varpi$ linearly dependent on wave phase and all other quantities constant. It is thus straightforward to average Equation \ref{lagrangian in limit}. Thus, the Whitham averaged Lagrangian density for a tightly wound eccentric wave with self-gravity and pressure is

\begin{equation}
\mathcal{L}_a = \langle L_a \rangle = - \frac{M_a n a^2}{N^2} \varphi_t \frac{\tilde{e} q}{2 k^2} + H^{\circ}_a \left( \frac{1}{2} \frac{\tilde{e}^2}{k^2} - F (q) - \frac{2 \lambda_{\rm sg}}{\bar{Q}^{\circ}} F_{\rm sg} (q) \right)
\end{equation}
where angle brackets denote a Whitham average over the phase of the wave and we have made use of $\tilde{\varpi}_t \tilde{e} = \varphi_t \tilde{\varpi}_{\varphi} \tilde{e} = \varphi_t q$. 

The dispersion relation is obtained by varying the Lagrangian with respect to the wave amplitudes $\tilde{e}$, $q$:

\begin{equation}
\frac{\delta \mathcal{L}}{\delta \tilde{e}} = 0 , \quad \frac{\delta \mathcal{L}}{\delta q} = 0 \quad, 
\end{equation}
combined with the compatibility relation

\begin{equation}
N \partial_t k = a \partial_a \varphi_t \quad .
\end{equation}
At next order the evolution of the wave fluxes are given by

\begin{equation}
\frac{\delta \mathcal{L}}{\delta \varphi}  = 0 \quad .
\end{equation}
The averaged Lagrangian has the gauge symmetry

\begin{equation}
k \rightarrow \lambda (a,t) k \quad, \qquad \varphi_t \rightarrow \lambda (a,t) \varphi_t \quad, \qquad \tilde{e} \rightarrow \lambda (a,t) \tilde{e} 
\end{equation}
where $\lambda (a,t)$ is an arbitrary function. This relates to our freedom to redefine the wave phase owing to the fact $\tilde{e}$ and $q$ do not vary on the short length/timescale. We can remove this gauge freedom by setting $q = \tilde{e}$. Upon adopting this gauge the Lagrangian becomes

\begin{equation}
\mathcal{L}_a = - \frac{M_a n a^2}{N^2} \varphi_t \frac{q^2}{2 k^2} + H^{\circ}_a \left( \frac{1}{2} \frac{q^2}{k^2} - F (q)  - \frac{2 \lambda_{\rm sg}}{\bar{Q}^{\circ}} F_{\rm sg} (q) \right)  \quad,
\end{equation}
with the dispersion relation now just given by variation with $q$,

\begin{equation}
\frac{\delta \mathcal{L}}{\delta q} = 0 \quad,
\end{equation}
while the compatibility relation and variational equation for the wave flux evolution are the same. The dispersion relation is given by

\begin{equation}
0 = - \frac{M_a n a^2}{N^2} \varphi_t \frac{q^2}{k^2} + H^{\circ}_a \left( \frac{q^2}{k^2} - q F^{\prime} (q)  - \frac{2 \lambda_{\rm sg}}{\bar{Q}^{\circ}} q F^{\prime}_{\rm sg} (q) \right)  \quad .
\end{equation}
Substituting this into the compatibility relation we obtain an equation in conservative form:

\begin{equation}
\frac{\partial k}{\partial t}  = \frac{\partial}{\partial y} \Biggl\{ \left(  \omega_{f} + \omega_{\rm sg} - \omega \right) \left[ 1 - k^2 q^{-1} \left( F^{\prime} (q)  - \frac{2 \lambda_{\rm sg}}{\bar{Q}^{\circ}} F^{\prime}_{\rm sg} (q) \right) \right]  \Biggr \} ,
\label{ideal compatability eq}
\end{equation}
where $y = \int \frac{N}{a} d a$. The equation for the wave flux evolution is

\begin{equation}
0 = \partial_t \left( \frac{M_a n a^2}{N^2} \frac{q^2}{2 k^2} \right) + \partial_a \left\{\frac{a H^{\circ}_a }{N} \left[  \frac{q F^{\prime} (q)}{k} + \frac{2 \lambda_{\rm sg}}{\bar{Q}^{\circ}} \frac{q F^{\prime}_{\rm sg} (q)}{k} \right]  \right\} ,
\label{ideal flux eq}
\end{equation}
where we have made use of the dispersion relation to simplify the spatial component of the flux. Consider steady solutions to these equations. Without loss of generality we can set $\varphi_t = 0$ ($\varphi_t$ must be constant so that $\partial_t k = 0$ and can be set to zero by a suitable choice of rotating frame). Then the equations become

\begin{equation}
1 = k^2 q^{-1} F^{\prime} (q)  + \frac{2 \lambda_{\rm sg}}{\bar{Q}^{\circ}} k^2 q^{-1} F^{\prime}_{\rm sg} (q)
\label{steady dispersion}
\end{equation}

\begin{equation}
\frac{a H^{\circ}_a }{N} \frac{q}{k} \left[  F^{\prime} (q) + \frac{2 \lambda_{\rm sg}}{\bar{Q}^{\circ}} F^{\prime}_{\rm sg} \right] = \mathrm{const} \quad .
\label{steady flux}
\end{equation}
Here Equation \ref{steady dispersion} is an amplitude dependent dispersion relation, while Equation \ref{steady flux} describes the conservation of angular momentum flux in the disc. When combined these equations determine how $k$ and $q$ vary with $a$ for a given disc profile. 

\begin{figure}
\centering
\includegraphics[trim=20 0 20 30,clip,width=\linewidth]{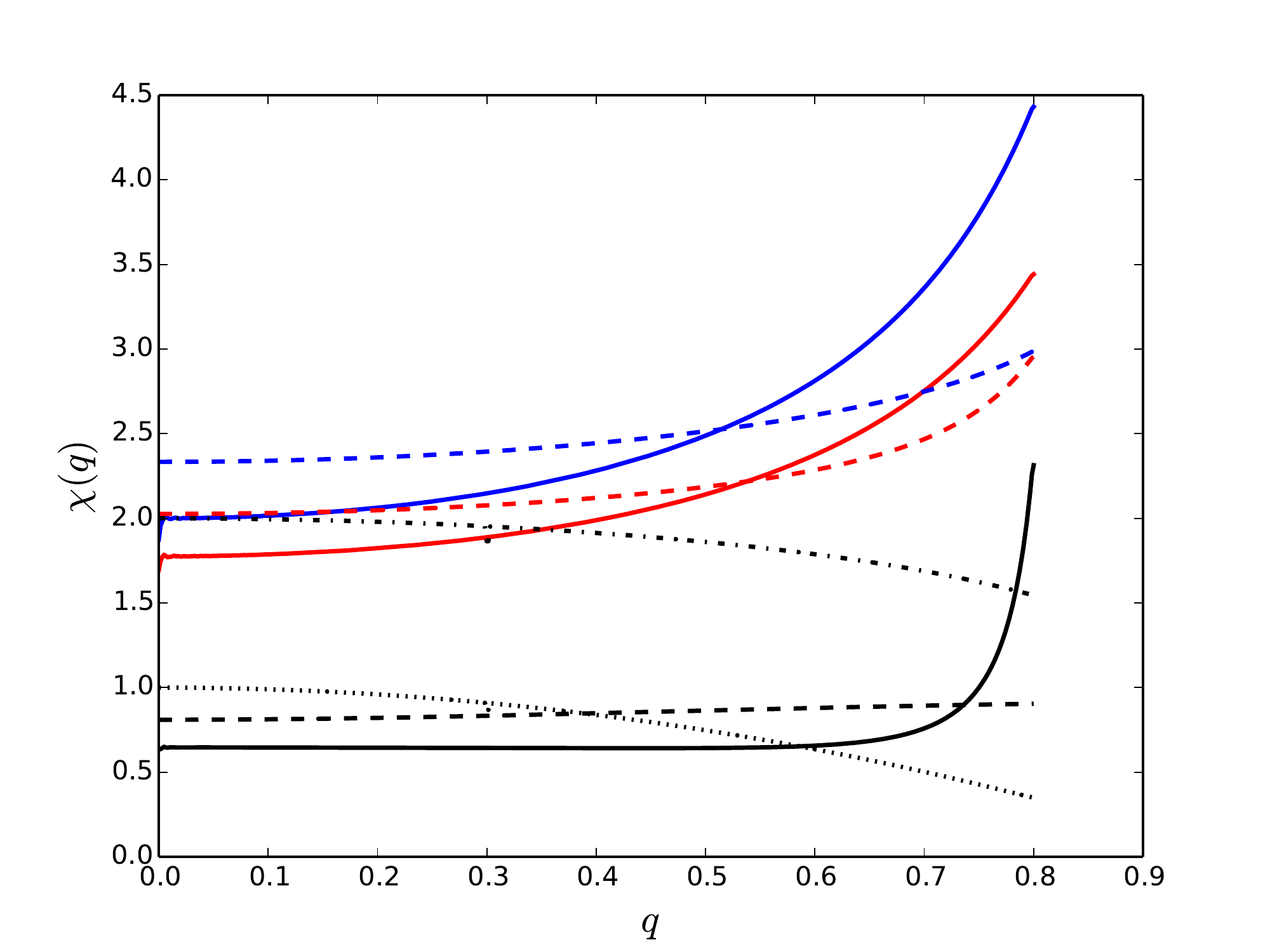}
\caption{Dependence of the nonlinear self-gravity control parameter on $q$ for different equations of state. Solid lines are isothermal (strictly $\gamma=1.00000001$ for numerical reasons) and dashed lines are $\gamma=2$. Colours indicate $\bar{Q}^{\circ}$, with black lines having $\bar{Q}^{\circ} = 10$, red lines having $\bar{Q}^{\circ} = 100$ and blue lines having $\bar{Q}^{\circ} = 10^8$. The dotted line is the equivalent calculation in a 2D disc with $\gamma=2$, while the dash-dotted line is an isothermal 2D disc.} 
\label{control param}
\end{figure}

Rearranging the dispersion relation for $k$,

\begin{equation}
k^2 = \frac{q}{F^{\prime} (q)} \left(1 + \frac{2 \lambda_{\rm sg}}{\bar{Q}^{\circ}} \chi (q)\right)^{-1}\quad,
\end{equation}
where we have introduced a self-gravity nonlinear control parameter,

\begin{equation}
\chi(q) = \frac{F^{\prime}_{\rm sg} (q)}{F^{\prime} (q) }  \quad,
\end{equation}
which determines whether pressure or self gravity becomes more important as the wave becomes nonlinear. The relative importance of self gravity to pressure forces is then set by a combination of the Toomre-Q in the reference disc and this nonlinear control parameter $\chi(q)$. Figure \ref{control param} shows $\chi(q)$ for a variety of disc models and equations of state. In 3D discs self gravity becomes more important as the wave becomes increasingly nonlinear. This contrast with 2D discs where $\chi(q)$ (calculated using the theory presented in \citet{Lee99}) decreases with increasing $q$, indicating that self-gravity becomes less important as the wave becomes more nonlinear. This strengthening of the disc self-gravity, relative to pressure, can potentially be explained by the fact that the wavelength of the nonlinear wave is longer than that predicted by linear theory as a result of strong pressure gradients preventing an orbital intersection which would form if the wavelength were too short.

Another important feature of self-gravitating discs in the tight-winding limit is that isothermal discs exhibit a breathing-mode excited by the disc self-gravity. This arise from the local contribution to $\psi_{\rm sg}$ which is proportional to $j^{-1}$, introducing a dependence on the lateral fluid compression. This means 3D, isothermal, self-gravitating discs in the tight-winding limit have different dynamics to 2D discs. This contrasts with the non self-gravitating discs where the disc's vertical structure doesn't affect the eccentric waves in the short-wavelength limit. More generally the modification of the disc scale height by self-gravity leads to $F(q)$ and $F_{\rm sg} (q)$ (along with similar functions describing viscosity which will be introduced later) depending on $\bar{Q}^{\circ}$.

Following \citet{Lee99} we can rewrite Equations \ref{ideal compatability eq} and \ref{ideal flux eq} in hyperbolic form,

\begin{equation}
 \frac{\partial}{\partial t} \begin{pmatrix}
 k \\
 j_t \\
 \end{pmatrix} + \mathbf{A} \frac{\partial}{\partial a} \begin{pmatrix}
 k \\
 j_t \\
 \end{pmatrix} = \begin{pmatrix}
 \frac{a}{N} \frac{\partial^2 \mathcal{L}_a}{\partial j_t \partial a} \\
 \frac{a}{N} \frac{\partial^2 \mathcal{L}_a}{\partial j_t \partial a} + \frac{\partial}{\partial a} \left( \frac{a}{N} \right) \frac{\partial \mathcal{L}_a}{\partial j_t} \\
 \end{pmatrix} \quad ,
\end{equation}
where $j^t = \frac{M_a n a^2}{N^2} \frac{q^2}{2 k^2}$ and the matrix $\mathbf{A}$ is given by

\begin{equation}
 \mathbf{A} = -\frac{a}{N} \begin{pmatrix}
 \frac{\partial^2 \mathcal{L}_a}{\partial j^{t} \partial k} & \frac{\partial^2 \mathcal{L}_a}{\partial (j^{t})^2} \\
 \frac{\partial^2 \mathcal{L}_a}{ \partial k^2} & \frac{\partial^2 \mathcal{L}_a}{\partial j^{t} \partial k} \\
 \end{pmatrix} \quad .
\end{equation}
This matrix has eigenvalues $V_{\pm}$, corresponding to the characteristic velocities, of

\begin{align}
 \begin{split}
 V_{\pm} =& \frac{N^2 \mathcal{H}_a^{\circ}}{M_a n a^2} \frac{N k}{a q} \Biggl [ F_{\rm p+sg}^{\prime} (q) + q F_{\rm p+sg}^{\prime \prime} (q) \\
  &\pm \sqrt{q F_{\rm p+sg}^{\prime \prime} (q) (q F_{\rm p+sg}^{\prime \prime} (q) - F_{\rm p+sg}^{\prime} (q))} \Biggr] \quad ,
 \end{split}
\end{align}
where $F_{\rm p+sg} (q) = F (q) + \frac{2 \lambda_{\rm sg}}{\bar{Q}^{\circ}} F_{\rm sg} (q)$. This system of equations are hyperbolic provided that $V_{\pm}$ are real, which is the case if $F^{\prime}_{\rm p+sg} (q)$ is a convex function of $q$ (As expected for most disc models of interest). As found in \citet{Lee99}, the sign of the characteristic velocities $\sgn (V_{\pm}) = \sgn(k)$. Thus these characteristic velocities are negative for tightly-wound trailing waves (which have $k < 0$).

\section{Modulation of the eccentric wave on the disc length scale} \label{modulation}

In this section we neglect self gravity and consider how the eccentricity and nonlinearity vary on the long-lengthscale, so that we can compare tightly wound waves with the untwisted discs considered in Paper I. We shall show the conditions for wave steepening derived in Paper I carry over to tightly wound waves. For comparison with Paper I it is useful to introduce $I (q)$ and $J(q)$ which for the tightly wound waves considered in this paper are given by

\begin{equation}
I (q ) := \frac{\pi}{\sqrt{2}} \sqrt{\frac{F^{\prime} (q)}{q}} \quad,
\end{equation}

\begin{equation}
J (q) := \frac{\pi}{2 \sqrt{2}} q^{1/2} F^{\prime} (q)^{3/2} \quad .
\end{equation}
These are directly comparable to the functionals $I(q)$, $J(q)$ defined in Paper I. Although we omit a proof here, one can show that if $F^{\prime} (q)$ is strictly convex on the interval $0< q < 1$ (a similar, but not identical, assumption to that made in Paper I) then $I(q)$ and $J(q)$ are increasing functions of $q$ on this interval. For isothermal and 2D adiabatic discs $1 < \gamma \le 2$ one can prove that $F(q)$ has the desired property, while for 3D adiabatic discs one can demonstrate these properties numerically.

\subsection{Flux conservation and nonlinear dispersion relation in a steady eccentric disc}

We shall now show that the wave steepening criteria derived in Paper I also apply to steady tightly-wound waves, in the absence of self-gravity. The dispersion relation for a steady, tightly wound, wave in the absence of self gravity is

\begin{equation}
\frac{\delta \langle L \rangle}{\delta q} = H_a^{\circ} F^{\prime} (q) \left( \frac{\pi^2}{2 k^2 I^2 (q)} - 1 \right) = 0 \quad.
\end{equation}
After rearranging and rescaling, this becomes

\begin{equation}
a \varphi_a = \frac{\pi N}{\sqrt{2} I (q)} \quad .
\label{dispersion relation llscl}
\end{equation}
The flux conservation law is

\begin{equation}
\frac{\delta \langle L \rangle}{\delta \varphi} = - \frac{\partial}{\partial a} \left( \frac{a H^{\circ}_a}{N} \frac{\partial \langle \hat{\mathcal{L}} \rangle}{\partial k}\right) = - \frac{\partial}{\partial a} \left( \frac{2 \sqrt{2} a H^{\circ}_a}{\pi N} J (q)\right) =0 ,
\end{equation}
which can be rearranged to obtain a conserved flux,

\begin{equation}
\frac{2 \sqrt{2} a H^{\circ}_a}{\pi N} J (q) = \mathrm{const} \quad .
\label{conserved flux untwisted}
\end{equation}

By writing $J(q)$ in terms of $I(q)$ and $q$, and making use of the gauge fixing condition $q = \tilde{e} = e N k$, we can obtain the alternative form of the flux conservation equation, which matches that derived in paper I for untwisted eccentric waves,

\begin{equation}
\frac{a H_{a}^{\circ} N^2}{a \varphi_a} \langle e^2 \rangle= \mathrm{const} \quad .
\end{equation}
Combining this with the dispersion relation we obtain an equation for the eccentricity envelope,

\begin{equation}
\langle e^2 \rangle \propto \frac{1}{a H_{a}^{\circ} N I (q)} \quad .
\label{untwisted amplitude}
\end{equation}

In the linear limit (using $F = F^{\rm lin}$, from Paper I) we can evaluate $I$ and $J$ and obtain

\begin{equation}
I (q)  = \frac{\pi}{2} \sqrt{\frac{2 \gamma - 1}{\gamma}} \quad ,
\end{equation}

\begin{equation}
J (q) = \frac{\pi}{8} \left(\frac{2 \gamma - 1}{\gamma} \right)^{3/2} \langle q^{2} \rangle \quad .
\end{equation}
This is identical to those for the untwisted eccentric mode, and matches the expressions given in Paper I upon the substitution $\langle q^2 \rangle = \frac{1}{2} q_{+}^2$ . 

In Paper I we showed that untwisted eccentric waves can be focused by the large scale disc properties and we derived conditions for $q$ and $e$ to become nonlinear in the inner disc from an initially linear disturbance in the outer disc. Here we show that the conditions derived in Paper I apply to tightly-wound waves.
 
As noted above, $I (q)$, $J (q)$ are increasing functions of $q$ for physically reasonable $F(q)$. Using Equation \ref{conserved flux untwisted} we can obtain the condition on the amplification of $q$ with decreasing $a$,

\begin{equation}
 \frac{\partial q}{\partial a} < 0 \quad \textrm{if and only if} \quad \frac{\partial}{\partial a} \left( \frac{a H^{\circ}_a}{N} \right) > 0 .
\end{equation}

\begin{figure}
\centering
\includegraphics[trim=40 10 40 50,clip,width=0.8\linewidth]{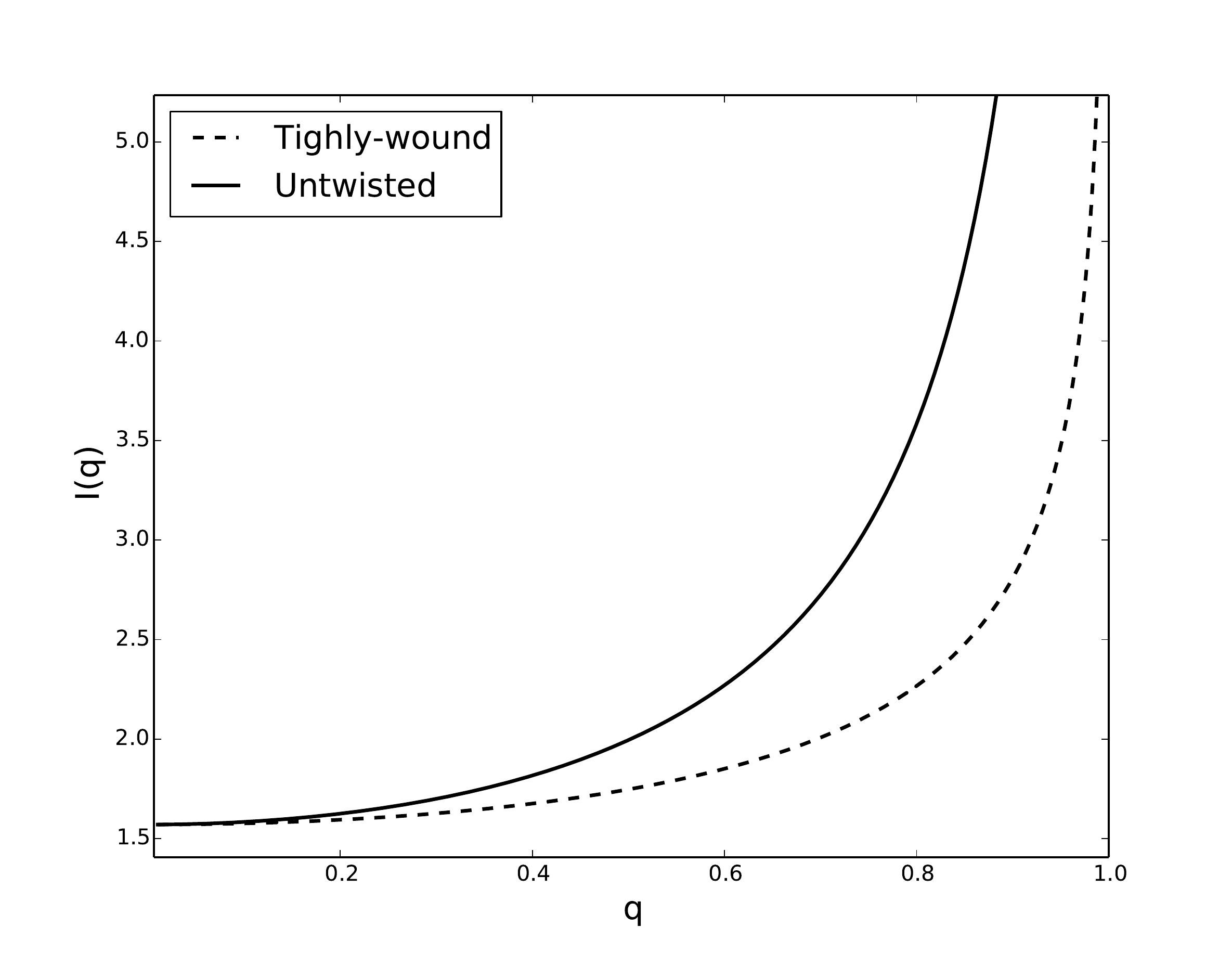}
\caption{Comparison of $I(q)$ for untwisted and tightly wound waves in an isothermal disc showing the initial weak dependence on $q$ for the tightly-wound waves.} 
\label{I comparison}
\end{figure}

The condition on $\langle e^2 \rangle$ to increase can be obtained from Equation \ref{untwisted amplitude}, from which we obtain the condition

\begin{equation}
\frac{\partial}{\partial a} \left( a H^{\circ}_a N \langle e^2 \rangle \right) < 0 \quad  \textrm{if and only if}  \quad \frac{\partial q}{\partial a} > 0 \quad .
\end{equation}
Similarly we have a condition on $a \varphi_a$:

\begin{equation}
\frac{\partial}{\partial a}  \left( \frac{a \varphi_a}{N} \right) > 0 \quad \textrm{if and only if}  \quad \frac{\partial q}{\partial a} > 0 \quad .
\end{equation}
These conditions are identical to those derived in Paper I. As done in Paper I, we can use these conditions to show that

\begin{equation}
\frac{a \varphi_a}{a \varphi_a |_{\rm lin}} = \frac{I(0)}{I(q)} \le 1 \quad ,
\end{equation}

\begin{equation}
\frac{\langle e^2 \rangle}{\langle e^2_{\rm lin} \rangle} = \frac{I(0)}{I(q)} \le 1 \quad ,
\label{nonlin e lin e}
\end{equation}
where the subscript $_{\rm lin}$ denotes the prediction from linear theory. The interpretation of this result is the enhancement of the pressure forces by the nonlinearity which can more easily balance the strong precessional forces, compared with linear theory. $I(q)$ for tightly wound waves often has a weaker dependence on $q$ than $I(q)$ of the untwisted theory and for an isothermal disc $\langle e^2 \rangle$ and $a \varphi_a$ remain close to their linear values until $q$ gets close to $1$ (See Figure \ref{I comparison}).

\section{Non-conservative terms} \label{pseudo lagrangian prescription}

In order to include non-conservative terms in our theory we need to move beyond the Lagrangian framework. To do this we make use of a pseudo-Lagrangian framework and the extension of the Whitham averaging method to pseudo-Lagrangians by \citet{Jimenez76}. In their formalism, non-conservative terms are treated perturbatively, and do not influence the dynamics over a single wave-cycle. These non-conservative terms instead affect the long time and length-scale modulation of the wave.

In the pseudo-Lagrangian formalism one doubles up the phase space variables and make the pseudo-Lagrangian stationary with respect to one set of phase space variables, whilst the `complementary' phase space variables are held fixed. There is an unfortunate notation clash between the notation introduced in \citet{Jimenez76} for complementary phase space variables (i.e. $\cdot^{\prime}$) and the notation for a derivative for a function of one variable, which we have made extensive use of to denote $q$ derivatives of the geometric part of the Hamiltonian $F$. To avoid this we shall adopt the convention that complementary phase space variables are denoted by a breve ($\breve{\cdot}$).

\subsection{The averaged pseudo-Lagrangian formalism}

As pseudo-Lagrangians and the formalism of \citet{Jimenez76} are not commonly used in astrophysics we shall use a damped linear wave equation, with a quadratic potential, as an illustrative example of how the formalism works. The damped linear wave equation, with quadratic potential, can be derived from the following pseudo-Lagrangian,

\begin{equation}
F_{\rm wave} = \int \left[ \frac{1}{2} u_t^2 - \frac{1}{2} c^2(x) u_x^2 - \frac{1}{2} m^2 (x) u^2 - D(x) u \breve{u}_t  \right] \, d x ,
\end{equation}
where $c$, $D$ and $m$ are position dependant wavespeed, damping rate and mass. Making the pseudo-Lagrangian stationary with respect to variations of $u$ and setting $\breve{u} = u$ we arrive at the equation of motion for the wave,

\begin{equation}
 \frac{\delta F_{\rm wave}}{\delta u} \Biggl |_{\breve{\cdot} = \cdot} = -u_{t t}+ \partial_x \left( c^2(x) u_x \right) - m^2 (x) u - D(x) u_t = 0 .
\end{equation}

When $c(x)$, $D(x)$ and $m(x)$ vary slowly over one wavelength of the wave we can make use of the  averaging formalism of \citet{Jimenez76}. Introducing the wave phase $\varphi$ with wavefrequency $\omega = \varphi_t$ and wavenumber $k = \varphi_x$ we can approximate the derivatives as $\partial_t \approx \omega \partial_{\varphi}$ and $\partial_x \approx k \partial_{\varphi}$.

\begin{align}
\begin{split}
F_{\rm wave} = \int & \Biggl\{ \frac{1}{2} \left[\omega^2 - c^2 (x) k^2 \right] u_{\varphi}^2 - \frac{1}{2} m^2 (x) u^2 \\
& - D (x) \, u \, \breve{\omega} \, \breve{u}_{\breve{\varphi}}  \Biggr \} \, d x .
\end{split}
\end{align}
On the short lengthscale damping can be neglected and $c$ can be treated as a constant. The short lengthscale dynamics are just a linear harmonic oscillator and can be solved by $u = u_0 (t,x) \cos \varphi(x)$, where $u_0 (t,x)$ is a time and position dependant wave amplitude. For the non-conservative terms we can make use of the following approximation $u(\varphi) \approx \breve{u} (\varphi) \approx \breve{u}(\breve{\varphi}) + (\varphi - \breve{\varphi}) \breve{u}_{\breve{\varphi}} (\breve{\varphi})$, as higher order terms in this expansion will vanish when we set $\breve{u} = u$, $\breve{\varphi} = \varphi$.

Averaging over the wave phase we obtain an averaged pseudo-Lagrangian for the system,

\begin{equation}
\begin{split}
\langle F_{\rm wave} \rangle = & \int \Biggl\{ \frac{1}{4} \left[\omega^2 - c^2 (x) k^2 - m^2 (x) \right] u^2_0 \\
&- \frac{1}{2} D (x) (\varphi - \breve{\varphi}) \breve{\omega} \breve{u}^2_0 \Biggr\} \, d x ,
\end{split}
\end{equation}
where angle bracket denote a phase average and we have dropped a term proportional to $\breve{u} \breve{u}_{\breve{\varphi}}$ which has no influence on the dynamics. Varying $\langle F_{\rm wave} \rangle$ with respect to the amplitude, $u_0$ and setting equal to zero we obtain,

\begin{equation}
\frac{\delta \langle F_{\rm wave} \rangle}{\delta u_0} \Biggl |_{\breve{\cdot} = \cdot} =  \frac{1}{2} \left[\omega^2 - c^2 (x) k^2 - m^2 (x) \right] u_0 = 0,
\end{equation}
from which we obtain the dispersion relation $\omega^2 = c^2 (x) k^2 + m^2 (x) $. Combined with the compatibility relation $\partial_t k = \partial_x \omega$, these determine how the wavephase is modulated on the long length/time scale. Varying with respect to $\varphi$, and setting equal to zero, we obtain an evolutionary equation for the wave amplitude,

\begin{align}
\begin{split}
\frac{\delta \langle F_{\rm wave} \rangle}{\delta \varphi} \Biggl |_{\breve{\cdot} = \cdot} &= -\frac{\partial}{\partial t} \left( \frac{1}{2}  \omega u^2_0 \right) + \frac{\partial}{\partial x} \left( \frac{1}{2} c^2 (x) k u^2_0 \right)  - \frac{1}{2} D (x) \omega u^2_0 \\
&= 0 .
\end{split}
\end{align}
This determines how the amplitude is modulated by the spacial dependence of $c$, $D$ and $m$. This can be rearranged into to obtain an equation of the form,

\begin{equation}
\nabla_{\mu} j^{\mu} = - D (x) j^t ,
\end{equation}
where $\mu = \{0,1\}$ are space time indices and $j^{\mu}$ are the wave flux. This is a waveflux conservation equation with a sink due to wavedamping.

\subsection{Viscous stresses}

In this section we consider the effects of shear and bulk viscous stresses on tightly wound waves. Viscous stresses are often used as a model for the effects of disc turbulence. The classical $\alpha$-disc model only considers shear viscosity, as a circular Keplerian flow is not affected by bulk viscosity. Turbulence due to the parametric instability feeds off the eccentric mode and has been shown to cause eccentricity to decay \citep{Wienkers18,Papaloizou05b}, it thus behaves like an effective bulk viscosity. The finite response time of the MRI can also act as an effective bulk viscosity that is typically of similar magnitude to the shear viscosity \citep{Ogilvie01,Lynch21}. Thus, the relative importance of shear and bulk viscosity depends on the source of disc turbulence, but one expects that they should be of broadly comparable magnitude.

We assume the vertically integrated shear viscosity $\bar{\mu}$ and bulk viscosity $\bar{\mu}_b$ can be written in the form $\bar{\mu} = \frac{\alpha \mu_a^{\circ}}{2 \pi} m$ and $\bar{\mu}_b = \frac{\alpha_b \mu_a^{\circ}}{2 \pi} m$, where $\alpha$, $\alpha_b$ are dimensionless constants, $\mu_a^{\circ}$ depends only on $a$ and $m$ is a function of $j$ and $h$ only. $\frac{\alpha \mu_a^{\circ}}{2 \pi}$ and $\frac{\alpha_b \mu_a^{\circ}}{2 \pi}$ are the vertically integrated shear and bulk viscosities of the reference circular disc, while $m$ is a dimensionless function of the disc geometry. As $j = j(q,\cos \tilde{E})$ and $h = h(q,\cos \tilde{E})$ we can write $m = m(q,\cos \tilde{E})$.

In the $\tilde{e} = q$ gauge, dissipative effects from viscous stresses in the tight-winding limit can be included using a pseudo-Lagrangian of the following form

\begin{align}
\begin{split}
F &= \frac{2}{3} \alpha \int a T_a^{\circ} (\varphi - \breve{\varphi}) [ F_{\rm shear} (\breve{q}) + 3 F_{\rm phase} (\breve{q}) ]\, d a \\
&+ \left( \alpha_b - \frac{2}{3} \alpha \right) \int a T_a^{\circ} (\varphi - \breve{\varphi}) [F_{\rm bulk} (\breve{q}) - F_{\rm phase, b} (\breve{q}) ] \, d a \quad ,
\end{split}
\end{align}
which is derived in Appendix \ref{viscous damping and phase}. Here we have introduced $T_a^{\circ} = n \mu^{\circ}_a$ and $F_{\rm bulk}$, $F_{\rm shear}$, $F_{\rm phase \, b}$ and $F_{\rm phase}$ are nonlinear functions of $q$ (similar to $F$ and $F_{\rm sg}$) given in Appendix \ref{viscous damping and phase}. $F_{\rm bulk}$ and $F_{\rm shear}$ are the contribution from the horizontal viscous stresses while $F_{\rm phase}$ and $F_{\rm phase, b}$ are contributions from the phase shift, between the vertical and horizontal motion, induced by the viscosity. 
 
 Assuming that the viscosity can be written as a powerlaw in density, pressure and scale height then we have a functional form form for $m$,
 
 \begin{equation}
 m = j^x h^y \quad ,
 \end{equation} 
where for a standard $\alpha-$viscosity $x = - \gamma$, $y = - (\gamma - 1)$. In the linear limit ($q \ll 1$), $F_{\rm bulk}$ and $F_{\rm shear}$ take the following forms for a 2D disc:

\begin{equation}
F^{(2D)}_{\rm bulk} (q) = \frac{q^2}{2} + O(q^4) \quad ,
\end{equation} 

\begin{equation}
F^{(2D)}_{\rm shear} (q) = \frac{9}{4} q^2 \left(1 + x \right) + O(q^4) \quad.
\end{equation} 
In the 2D disc there is no phase shift so $F_{\rm phase} (q) =  F_{\rm phase \, b} (q) = 0$. In a 3D disc,

\begin{equation}
F^{(3D)}_{\rm bulk} (q) = \frac{1}{2 \gamma} q^2 + O(q^4) \quad,
\end{equation}

\begin{equation}
F^{(3D)}_{\rm shear} (q) = \frac{9}{4} q^2 \left(1 + x - y \frac{\gamma - 1}{\gamma} \right) + O(q^4) \quad ,
\end{equation} 

\begin{equation}
F_{\rm phase \ , b} = \frac{(\gamma - 1)}{2 \gamma^2} q^2 + O(q^4) \quad ,
\end{equation}

\begin{equation}
F_{\rm phase} =\frac{1}{2} \left( \frac{\gamma - 1}{\gamma} \right)^2 q^2 + O(q^4) \quad .
\end{equation}

\begin{figure}
\centering
\includegraphics[trim=0 10 0 25,clip,width=0.9\linewidth]{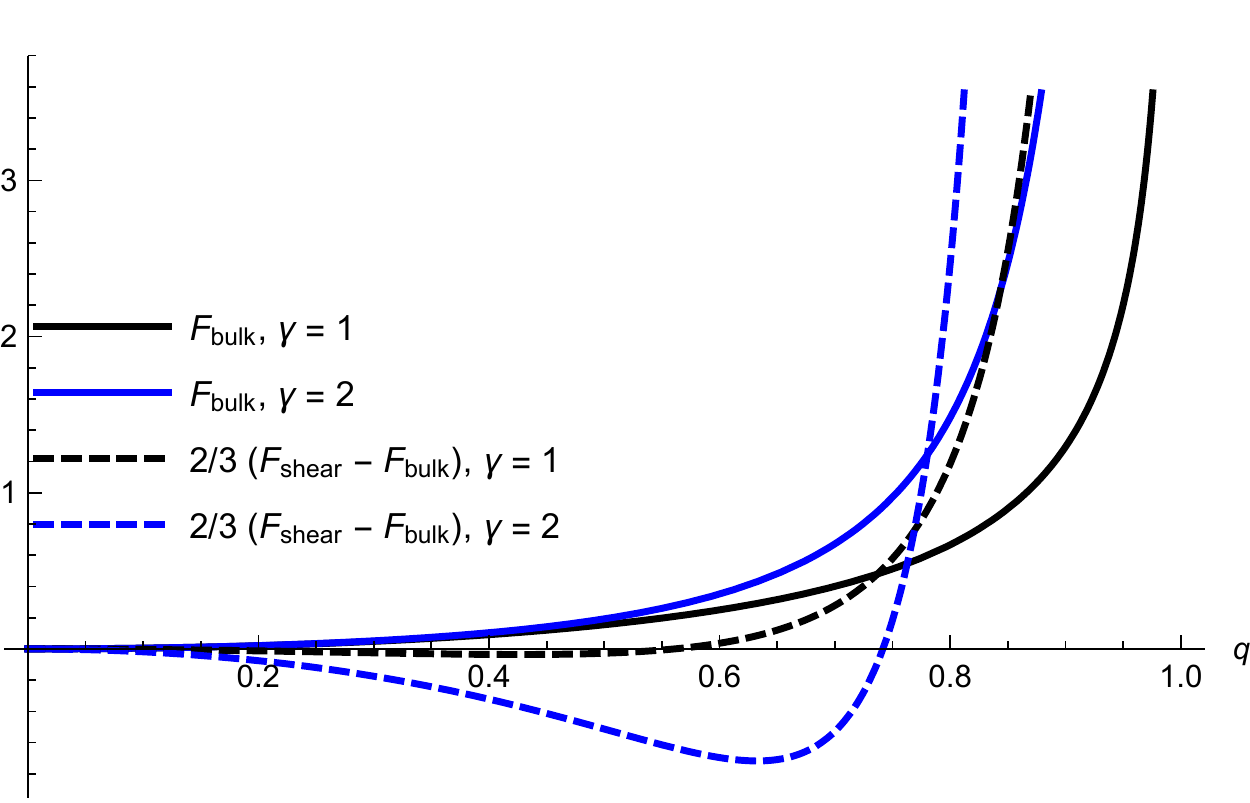}
\caption[Viscous integrals for different disc models]{Viscous integrals for a 2D $\gamma=1$ and $\gamma=2$ discs. Negative values indicates overstable behaviour. The shear viscosity initially results in overstability, but damps the wave at a critical $q$ which depends on the disc model. The same qualitative behaviour has long been known in planetary rings \citep{Borderies85,Borderies86}.} 
\label{viscous q dependance}
\end{figure}

Neglecting the phase shift, we require $F_{\rm shear} > F_{\rm bulk}$ in order that shear viscosity damps the wave. For an $\alpha-$disc, with $1 \le \gamma \le 2$, both $F^{(2D)}_{\rm shear} < F^{(2D)}_{\rm bulk}$ and $F^{(3D)}_{\rm shear} < F^{(3D)}_{\rm bulk}$, in the linear limit, so shear viscosity excites tightly-wound waves in the linear limit, recovering the known viscous overstability \citep{Syer92,Syer93,Lyubarskij94,Ogilvie01,Latter06}. When the wave is sufficiently nonlinear, shear viscosity instead damps the wave (See Figure \ref{viscous q dependance}). More generally in order that viscosity behaves dissipatively we require

\begin{align}
\begin{split}
\mathcal{D} &= \frac{2}{3} \alpha [F_{\rm shear} (q) + 3 F_{\rm phase} (q) -  F_{\rm bulk} (q) + F_{\rm phase \, b} (q) ] \\
&+ \alpha_b [F_{\rm bulk} (q) - F_{\rm phase \, b} (q) ]  > 0 .
\end{split}
\end{align}
In the linear limit, for a 2D disc this condition becomes

\begin{equation}
\mathcal{D}  \approx \frac{q^2}{6} \alpha  \left[  7 + 3 \frac{\alpha_b}{\alpha}  + 9 x \right] > 0 \quad ,
\end{equation}
while for a 3D disc,

\begin{equation}
\mathcal{D} \approx \frac{\alpha}{6} q^{2} \left[ 15 + 9 x - 9 y \frac{\gamma - 1}{\gamma} - \frac{4}{\gamma^2} (3 \gamma - 1) + \frac{\alpha_b}{\alpha} \frac{3}{\gamma^2} \right] > 0 .
\end{equation}
For both 2D $\alpha-$discs, and a 3D $\alpha-$disc (if the contribution from the phase shift is neglected), the viscous overstability is suppressed when

\begin{equation}
\frac{\alpha_b}{\alpha}  > \frac{9 \gamma - 7}{3} \quad .
\end{equation}
For a 3D $\alpha-$disc, including phase shift, the viscous overstability is suppressed when

\begin{equation}
\frac{\alpha_b}{\alpha}  > \frac{3 \gamma^2 + 3 \gamma -4}{3} .
\end{equation}
These conditions agree with the small $\alpha$, $\alpha_b$ limit of \citet{Latter06}, who similarly found that the phase shift in 3D discs made them more susceptible to the viscous overstability.

Viscosity in the disc also drives an accretion flow. While we do not directly include accretion in our model, as it typically occurs on a timescale longer than the secular timescale associated with the eccentric wave evolution, it is illustrative to consider the effects of the eccentric wave on the accretion flow. From Appendix \ref{eccentric accretion flow} the viscous torque, responsible for driving the accretion flow, in a tightly-wound eccentric disc is

\begin{equation}
 \mathcal{G} = - 2 \alpha a^2 T_a^{\circ} \frac{1}{2 \pi} \int m j^{-1} \left(\frac{1}{4} - j \right)\, d E \quad .
\end{equation}
As an illustrative case consider an isothermal $\alpha$-disc, where $m = j^{-1}$, one can evaluate the orbit average to obtain

\begin{equation}
  \mathcal{G}_{\rm iso} =  \frac{3}{2} \alpha a^2 T_a^{\circ} \frac{1 - \frac{4}{3} q^2}{(1 - q^2)^{3/2}} \quad .
\end{equation}
Neglecting gradients in $q$, when $q \ge \sqrt{3/4}$ the modification due to the eccentric wave is sufficient to halt the accretion flow, which may lead to the truncation of the disc on the viscous time. When $q \ll 1$ the torque simplifies to

\begin{equation}
  \mathcal{G}_{\rm iso} = \frac{3}{2} \alpha a^2 T_a^{\circ} \left(1 + \frac{1}{6} q^2 \right) + O(q^4) \quad .
\end{equation}
Thus in the linear limit, where the viscous overstability operates, the presence of the eccentric wave enhances the accretion flow. A similar enhancement of disc accretion by eccentric discs was found by \citet{Syer92,Syer93}. This enhancement arises from the density dependence of the $\alpha$-viscosity.

\begin{figure}
\centering
\includegraphics[trim=0 0 0 0,clip, width=0.8\linewidth]{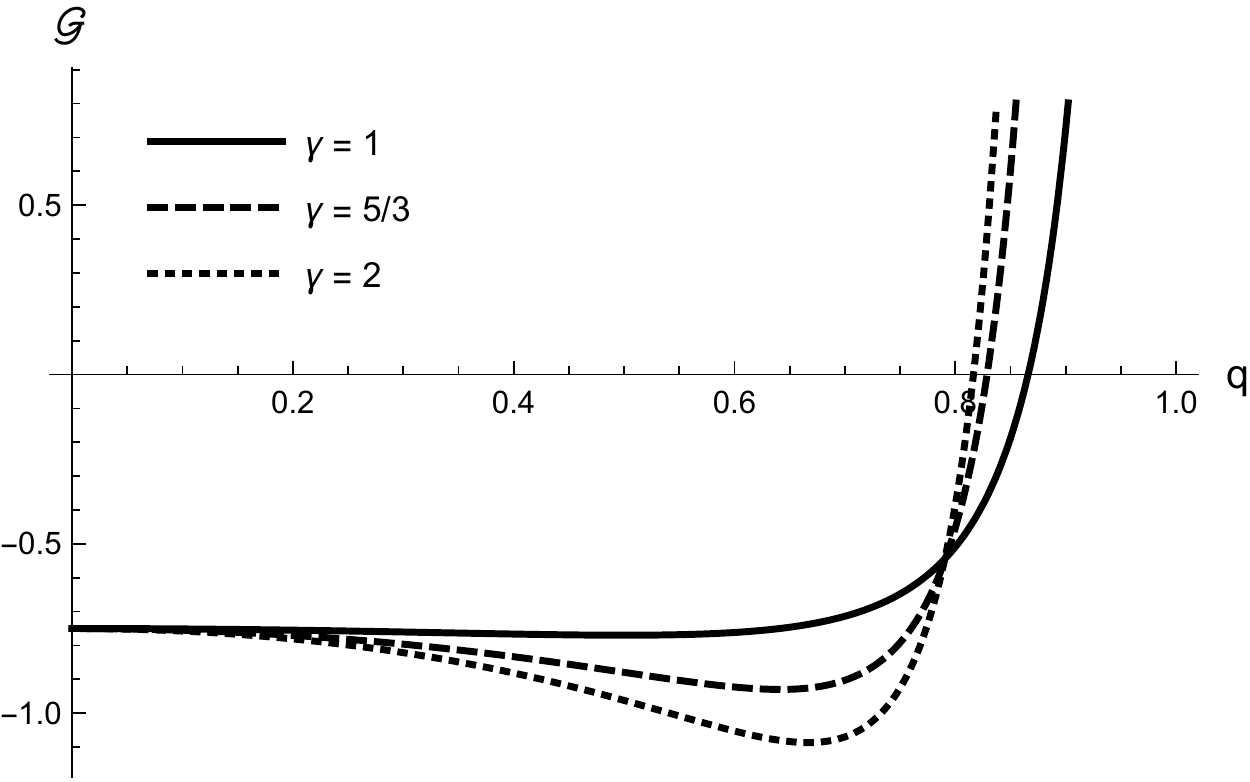}
\caption{Viscous torque for a tightly-wound eccentric disc, where $m = j^{-\gamma}$ corresponding to a 2D (3D for $\gamma = 1$) alpha disc. The torque is a strongly increasing function of $q$ for $q \sim 0.8$ and could halt the accretion flow if q increases inwards.} 
\label{viscous torque}
\end{figure}

Figure \ref{viscous torque} plots $\mathcal{G}$ against $q$ for different disc models. In general, weakly nonlinear waves drive (slightly) stronger accretion, while strongly nonlinear eccentric waves have the potential to significantly modify the accretion flow - including cutting it off entirely. As $\alpha$ is small by assumption, the modification of the background flow by the eccentric wave will be negligible on the timescale of interest except for strongly nonlinear eccentric waves. The presence of strongly nonlinear eccentric waves could lead to significant evolution of the background surface density profile on the timescale of interest, potentially resulting in disc truncation. 

The above analysis neglects gradients in $q$, which, when large enough, can fundamentally change the picture described above. This is particularly important for strongly nonlinear waves where $\mathcal{G}$ is a steep function of $q$. A more complete study of the effects of tightly-wound eccentric waves on the background accretion flow is beyond the scope of this paper, and is thus left for future work.

\section{Nonconservative, tight winding wave dynamics} \label{nonconservative wave dynamics}

The Whitham averaged pseudo-Lagrangian density for a tightly wound eccentric wave with self-gravity, pressure and viscosity, in the $\tilde{e} = q$ gauge, is:

\begin{align}
\begin{split}
\mathcal{F}_a &= -\frac{M_a n a^2}{N^2} \varphi_t \frac{q^2}{2 k^2} + H^{\circ}_a \left( \frac{1}{2} \frac{q^2}{k^2} - F (q) - \frac{2 \lambda_{\rm sg}}{\bar{Q}^{\circ}} F_{\rm sg} (q) \right)   \\
&+ \frac{2}{3} \alpha a T_a^{\circ} (\varphi - \breve{\varphi}) [F_{\rm shear} (\breve{q}) + 3 F_{\rm phase} (\breve{q})  \\
&- F_{\rm bulk} (\breve{q}) + F_{\rm phase b} (\breve{q})] \\
&+ \alpha_b a T_a^{\circ} (\varphi - \breve{\varphi}) [ F_{\rm bulk} (\breve{q})  - F_{\rm phase \, b} (\breve{q}) ] \quad .
\end{split}
\end{align}
The dispersion relation is obtained by varying the pseudo Lagrangian with respect to the wave amplitude $q$

\begin{equation}
\frac{\delta \mathcal{F}}{\delta q} \Biggr |_{\breve{\cdot}=\cdot} = 0 \quad,
\end{equation}
combined with the compatibility relation,

\begin{equation}
N \partial_t k = a \partial_a \varphi_t \quad .
\end{equation}
At next order the evolution of the wave fluxes are given by

\begin{equation}
\frac{\delta \mathcal{F}}{\delta \varphi} \Biggr |_{\breve{\cdot}=\cdot} = 0 \quad .
\end{equation}
The dispersion relation is given by

\begin{align}
\begin{split}
0 &= -\frac{M_a n a^2}{N^2} \varphi_t \frac{q^2}{k^2} + H^{\circ}_a \left[ \frac{q^2}{k^2} - q F^{\prime} (q) \left(1 + \frac{2 \lambda_{\rm sg}}{\bar{Q}^{\circ}} \chi (q)\right) \right] \quad ,
\end{split}
\label{dispersion full}
\end{align}
where we have made use of the self-gravity nonlinear control parameter $\chi (q)$. Substituting this into the compatibility relation we obtain an equation in conservative form,

\begin{equation}
 \partial_t k -  \partial_y \Biggl\{ (\omega_f + \omega_{\rm sg} - \omega) \left[ 1  - \frac{k^2}{q} F^{\prime} (q) \left(1 + \frac{2 \lambda_{\rm sg}}{\bar{Q}^{\circ}} \chi (q)\right)\right] \Biggr\} =0 \quad ,
\label{compatability full}
\end{equation}
where $y = \int \frac{N}{a} d a$. 

At the next order, the equation for the wave flux evolution is,

\begin{align}
\begin{split}
\partial_t & \left(\frac{M_a n a^2}{N^2} \frac{q^2}{2 k^2}\right) +  \partial_a \frac{a H^{\circ}_a }{N} \Biggl \{  \frac{q F^{\prime} (q)}{k} \left(1 + \frac{2 \lambda_{\rm sg}}{\bar{Q}^{\circ}} \chi (q)\right) \Biggr \} \\
&=- \frac{2}{3} \alpha a T_a^{\circ}  [F_{\rm shear} (\breve{q}) + 3 F_{\rm phase} (\breve{q})  - F_{\rm bulk} (\breve{q}) + F_{\rm phase b} (\breve{q})] \\
&- \alpha_b a T_a^{\circ} [ F_{\rm bulk} (\breve{q})  - F_{\rm phase \, b} (\breve{q}) ]  \quad ,
\end{split}
\label{amplitude full}
\end{align}
where we have made use of the dispersion relation to simplify the spatial component of the flux. 

Equations \ref{compatability full} and \ref{amplitude full} are evolutionary equations for the wavenumber $k$ and nonlinearity $q$. In the absence of viscosity these equations are conservative and describe how the wavenumber and angular momentum deficit (AMD, a positive definite measure of disc eccentricity, being the additional angular momentum that needs to be added to an orbit to make it circular) are conserved in the disc. In general this equation has the form

\begin{equation}
\nabla_{\mu} j^{\mu} = S ,
\end{equation}
where ${\mu}=\{0,1\}$ are space time indices, $j^t = \frac{M_a n a^2}{N^2} \frac{q^2}{2 k^2}$ is the AMD density $j^{a}$ is the AMD flux and $S$ are sources/sinks of AMD.

Now we confirm that the linear limit of our theory matches the WKB solutions of linear eccentric disc theory. In Paper I we showed this was the case for the ideal short-wavelength theory. For a 3D disc, the linearised dispersion relation, neglecting self-gravity\footnote{Self-Gravity can be important in the linear/WKB limit of our theory. This contrasts with the WKB limit of existing linear theory, where self-gravity is generally unimportant. This arises from considering the effect of the disc's vertical self-gravity, which is not normally accounted for.}, is

\begin{equation}
0 = -\frac{M_a n a^2}{N^2 k^2} \varphi_t + H^{\circ}_a \left( \frac{1}{k^2} - \frac{2 \gamma - 1}{2 \gamma} \right) \quad ,
\label{dispersion linear}
\end{equation}
while the equation for the wave flux evolution is

\begin{align}
\begin{split}
0 &= \partial_t \left(\frac{M_a n a^2}{N^2} \frac{q^2}{2 k^2} \right) + \partial_a  \Biggl( \frac{a H^{\circ}_a}{k N} \frac{q^2}{2}   \frac{2 \gamma - 1}{ \gamma}   \Biggr)  \\
&+  a T_a^{\circ} \Biggl \{ \frac{3}{2} \alpha \left[1 + x - y \frac{\gamma - 1}{\gamma} + \frac{2}{3} \left(\frac{\gamma - 1 }{\gamma} \right)^2 \right] \\
& + \left(\alpha_b - \frac{2}{3} \alpha \right) \frac{1}{2 \gamma^2} \Biggr\}  q^2 \quad .
\end{split}
\end{align}
Making use of $q = N k e$ we can rearrange this to obtain

\begin{align}
\begin{split}
0 &= M_a n a^2 e_t + a H^{\circ}_a  \Biggl( N k \frac{2 \gamma - 1}{ \gamma} \Biggr) e_a \\
&+ e \partial_a  \left[ \frac{a H^{\circ}_a}{2} \Biggl( N k \frac{2 \gamma - 1}{ \gamma} \Biggr) \right] \\
&+  a T_a^{\circ} \Biggl \{ \frac{3}{2} \alpha \left[1 + x - y \frac{\gamma - 1}{\gamma} + \frac{2}{3} \left(\frac{\gamma - 1 }{\gamma} \right)^2 \right]  \\
&+ \left(\alpha_b - \frac{2}{3} \alpha \right) \frac{1}{2 \gamma^2} \Biggr\} N^2 k^2 e \quad.
\end{split}
\label{e evolve linear}
\end{align}

From \citet{Teyssandier16} the dynamics of a linear eccentric disc subject to pressure and bulk viscosity is

\begin{align}
\begin{split}
\Sigma r^2 \Omega \dot{\mathcal{E}} &- i \Sigma r^2 \Omega \omega_{f} \mathcal{E} = \frac{i}{r} \frac{\partial}{\partial r} \left[ \frac{1}{2} \left( 2 - \frac{1}{\gamma}\right) P r^3 \frac{\partial \mathcal{E}}{\partial r} \right] \\
&+ \frac{i}{2} \left(4 - \frac{3}{\gamma}\right) r \frac{d P}{d r} \mathcal{E} + \frac{3}{2} i \left(1 + \frac{1}{\gamma}\right) P \mathcal{E} \\
&+ \frac{1}{2 r} \frac{\partial}{\partial r} \left( \alpha_b P r^3 \frac{\partial \mathcal{E}}{\partial r} \right) \quad .
\end{split}
\label{linear theory}
\end{align}
In the WKB limit terms we only keep terms involving derivatives of $E$, as well as the term involving the forced precession $\omega_f$ so the relevant equation is
 
\begin{align}
\begin{split}
\Sigma r^2 \Omega \dot{\mathcal{E}} - i \Sigma r^2 \Omega \omega_{f} \mathcal{E} &= \frac{i}{r} \frac{\partial}{\partial r} \left[ \frac{1}{2} \left( 2 - \frac{1}{\gamma}\right) P r^3 \frac{\partial \mathcal{E}}{\partial r} \right] \\
&+ \frac{1}{2 r} \frac{\partial}{\partial r} \left( \alpha_b P r^3 \frac{\partial \mathcal{E}}{\partial r} \right) \quad .
\end{split}
\label{wkb limit theory}
\end{align}
This scaling comes from assuming the precessional terms are large,  $\Sigma r^2 \Omega \omega_{\rm f} \gg P$. We also require the viscosity to be sufficiently small. The appropriate scaling to match the tight-winding theory is $\omega_{f} - \omega \sim \varpi_a^2 = O(\delta^{-2})$, $e_t \sim \varpi_a e$ and $\alpha_b = O(\delta)$. Splitting into real and imaginary parts and writing in terms of our notation Equation \ref{wkb limit theory} becomes,

\begin{align}
\begin{split}
 M_a n a^2 e_t &= - \varpi_a e \frac{\partial}{\partial a} \left( \frac{2 \gamma - 1}{2 \gamma}  H_{a}^{\circ} a^2 \right) \\
&- \left( \frac{2 \gamma - 1}{2 \gamma}  H_{a}^{\circ} a^2 \right) \left( e_a \varpi_a  + e \varpi_{a a} \right) \\
&- \frac{1}{2} \alpha_b T_a^{\circ} a^3 e \varpi_a^2 + \text{Higher order terms} ,
\end{split}
\label{Re wkb limit theory}
\end{align}

\begin{equation}
N^2 H_a^{\circ} e = \left( \frac{2 \gamma - 1}{2 \gamma} H_{a}^{\circ} a^2 \right) e \varpi_a^2 + \text{Higher order terms} .
\label{Im wkb limit theory}
\end{equation}
Noting that $a \varpi_a = a \varphi_a = k N$, Equations \ref{Re wkb limit theory} and \ref{Im wkb limit theory} match Equations \ref{e evolve linear} and \ref{e evolve linear} respectively (in the absence of self gravity and shear viscosity), with the exception of a constant factor in the viscous term. In fact Equation \ref{Re wkb limit theory} will match Equation \ref{e evolve linear} if the 2D form of the bulk viscosity ($F^{(2D)}_{\rm bulk}$) is used. In practice this difference can just be absorbed into the definition of $\alpha_b$.

\section{Tightly wound waves in an isothermal pseudo-Newtonian disc} \label{pseudo-Newtonian disc discussion}

\subsection{A comparison with \citet{Dewberry19b,Dewberry20}} \label{ideal comp with dewberry}

Recent simulations of eccentric discs in a pseudo-Newtonian disc were carried out by \citet{Dewberry19b} (hydrodynamic) and \citet{Dewberry20} (magnetohydrodynamic). In both papers an eccentric wave was excited at the outer boundary and propagated inwards, forming a twisted eccentric disc.  \citet{Dewberry19b,Dewberry20} considered a globally isothermal disc with a constant surface density and mimicked the the effects of GR through the Paczy\'{n}ski-Wiita potential \citep{Paczynsky80},

\begin{equation}
\Phi_{\rm PW} = -\frac{G M_1}{r - 2 r_g} \quad ,
\label{PW potential}
\end{equation}
where $r_g$ is the gravitational radius of the black hole.

One important finding of \citet{Dewberry20} was that while the eccentric wave became increasingly nonlinear as it approached the marginally stable orbit there was also a decrease in eccentricity. This effect was present in both hydrodynamic and magnetohydrodynamic runs and is thus unlikely to be related to turbulent dissipation of disc eccentricity. This is in contrast to the expectations of \citet{Ferreira09} where the eccentricity only declines as it approaches the marginally stable orbit if it is subject to sufficient damping.

Based on the theory presented in Paper I, \citet{Dewberry20} suggested that the decline in eccentricity was due to the nonlinear steepening of the wave as it approaches the marginally stable orbit, specifically the effect captured by equation \ref{nonlin e lin e}. In this section we shall show, that while true for untwisted disc, this mechanism doesn't work for tightly-wound (i.e. purely ingoing) waves. We shall show this effect can be explained either by a partial reflection of the wave or catastrophic damping from the presence of shear viscosity.

There is currently a lot of uncertainty in what boundary condition to apply at the marginally stable orbit. However, as the eccentric discs produced in \citet{Dewberry19b} and \citet{Dewberry20} are clearly twisted they are best compared with the tightly wound theory. The discs in these simulations are globally isothermal and are setup with a constant surface density exterior to the marginally stable orbit and set to a floor value interior to it. This relaxes to a background surface density which smoothly transitions from a constant value exterior to the marginally stable orbit to near zero interior to it (see Figure 3 of \citet{Dewberry19b}). To mimic this behaviour we adopt a mass per unit semimajor axis of

\begin{equation}
M_a = 2 \pi \Sigma^{\circ} a T(a) ,
\end{equation}
where $\Sigma^{\circ}$ is the constant surface density exterior to the marginally stable orbit and $T(a)$ is a smooth function that interpolates between 1 and 0 around the marginally stable orbit. The circular Hamiltonian density is given by,

\begin{equation}
H^{\circ}_a = 2 \pi c^2_0 \Sigma^{\circ} a T(a) ,
\end{equation}
with $c_0$ the constant sound speed.For numerical implementations we adopt a $\tanh$ function for $T(a)$ which is given by

\begin{equation}
T(a) = \frac{1}{2} \left[1 + \tanh \left(\frac{a - 2 r_g - 2 w_{\rm inner}}{w_{\rm inner}} \right) \right] ,
\end{equation}
where we adopt $w_{\rm inner} = 0.5$ to mimic the profiles shown in \citet{Dewberry19b} for $c_0 = 0.01$. 

For a Paczy\'{n}ski-Wiita potential the circular angular velocity is \citep{Paczynsky80}

\begin{equation}
\Omega = \sqrt{\frac{G M_1}{r^3}} \frac{r}{r - 2 r_g} 
\label{omega pw}
\end{equation}
and the epicyclic frequency is \citep{Okazaki87}

\begin{equation}
\kappa = \Omega \sqrt{\frac{r - 6 r_g}{r - 2 r_g}} \quad.
\label{kappa pw}
\end{equation}
Notably this is different from the exact epicyclic frequency measured at infinity which was adopted by \citet{Ferreira09}. 

As we are in the tight-winding limit, we can make the approximation $a \approx r$ and, using Equations \ref{omega pw} and \ref{kappa pw}, obtain the forced precession frequency from the Paczy\'{n}ski-Wiita potential,

\begin{equation}
\omega_f = \frac{2 n a^2 r_g}{(a - 2 r_g)^3} .
\end{equation}
Here we have used $n = \sqrt{\frac{G M_1}{a^3}}$ to denote the Keplerian mean motion. At large distances $\omega_f$ behaves like $\omega_f \propto a^{-5/2}$ which was used to model GR precession in the power law discs. This, along with surface density and pressure profiles, gives an expression for $N^2$,

\begin{equation}
N^2 = \frac{2 n^2 a^4 r_g}{c_0^2 (a - 2 r_g)^3} \quad ,
\end{equation}
which is independent of the disc truncation.

In the linear regime, $q$ and the shape of the eccentric envelope are given by

\begin{equation}
q \propto a^{-3/4} (a - 2 r_g)^{-3/4} T(a)^{-1/2} \quad,
\end{equation}

\begin{equation}
\langle e^2 \rangle \propto a^{-5/2} (a - 2 r_g)^{3/2} T(a)^{-1} \quad ,
\end{equation}
both of which diverge as $T (a) \rightarrow 0$ near the marginally stable orbit. It is worth noting that the singular behaviour here is not directly caused by GR precession, but is instead a consequence of the truncation of the disc (which is a consequence of the orbit stability in GR). A similarly truncated disc in the absence of GR precession would experience the same singular behaviour as $T (a) \rightarrow 0$. 

In the nonlinear theory we obtain an expression for $J(q)$ and $e$ in the disc,

\begin{equation}
J(q) \propto a^{-3/2} (a - 2 r_g)^{-3/2} T(a)^{-1} .
\end{equation}
Notably in the vicinity of the marginally stable orbit $T \rightarrow 0$ and $J(q)$ diverges, meaning $q \rightarrow 1$. Determining the behaviour of $e$ is less straightforward as while $e_{\rm lin} = (a H_a^{\circ} N)^{-1/2}$ diverges, so too does $I(q)$ as $q \rightarrow 1$. These two effects will compete against each other with the diverging $e_{\rm lin}$ causing $e$ to diverge, while the diverging $I(q)$ will cause $e$ to tend to zero. In the untwisted theory, with more realistic disc profiles, the nonlinearity causes the disc to circularise as it approaches the disc truncation at the marginally stable orbit. However nonlinearity has a weaker effect in tightly-wound discs, while the effect of the disc truncation on $e_{\rm lin}$, in the isothermal disc, is much stronger, so this result doesn't hold. 

\begin{figure}
\centering
\includegraphics[trim=0 0 0 0,clip,width=\linewidth]{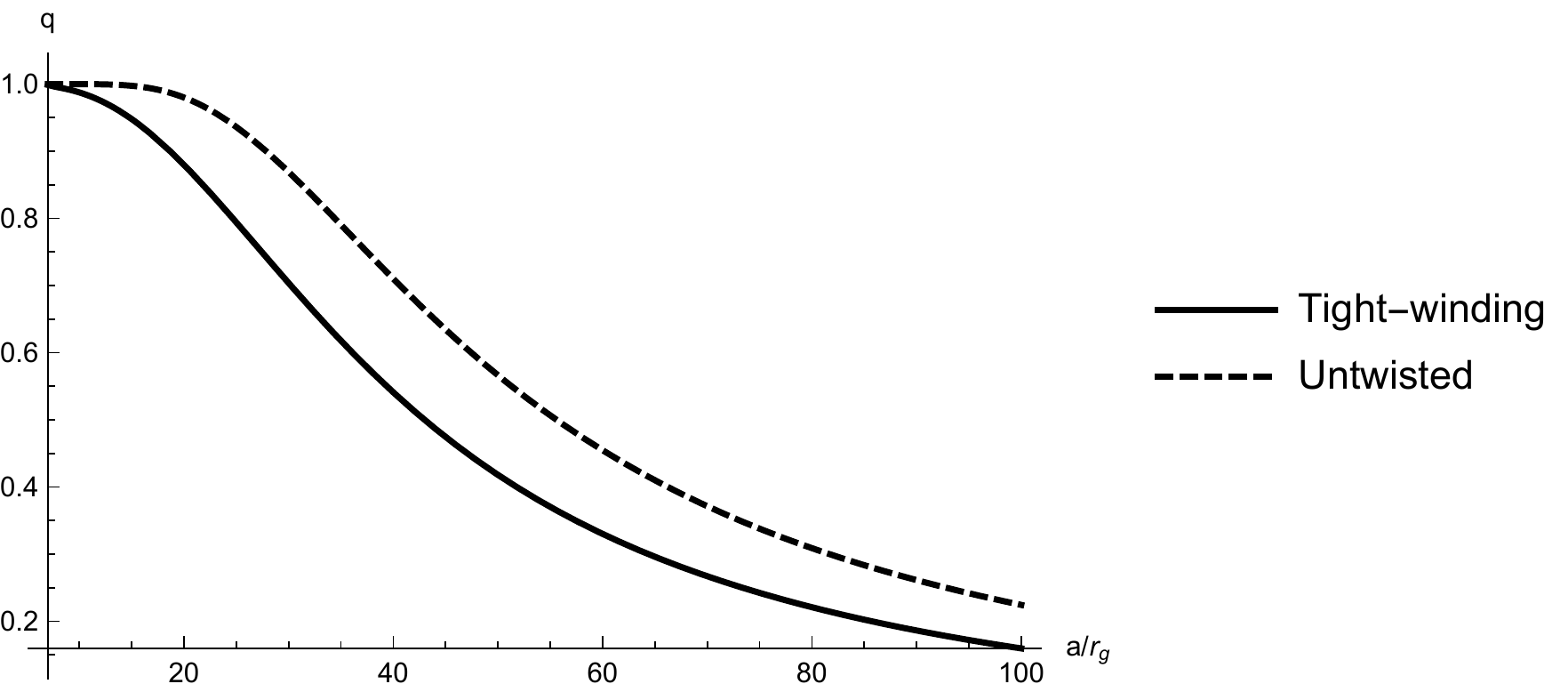}
\includegraphics[trim=0 0 0 0,clip,width=\linewidth]{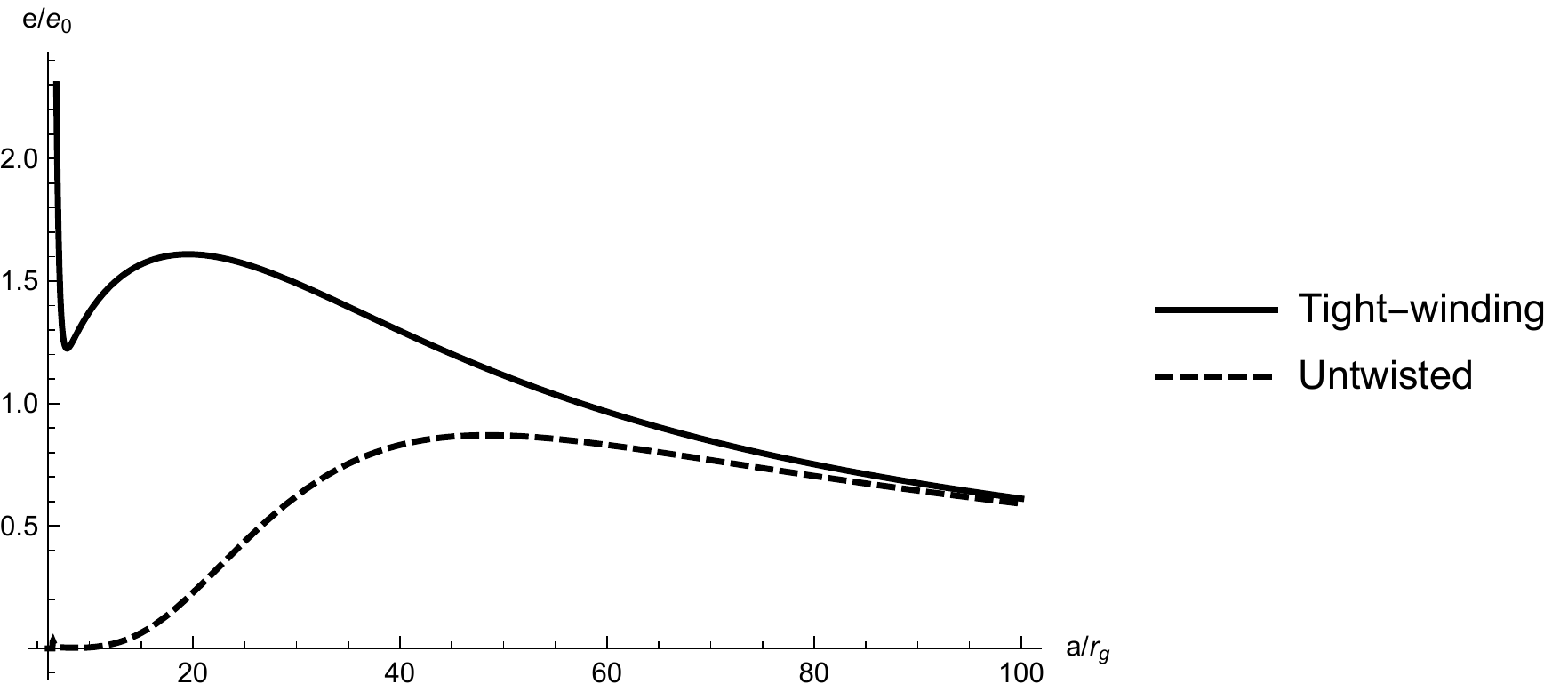}
\caption{Nonlinearity (top) and Eccentricity (bottom) for short-wavelength waves approaching the marginally stable orbit in an inviscid isothermal disc with truncation. The nonlinear effects in the tightly-wound wave (solid line) are too weak to regularise the singular behaviour associated with the disc truncation. The nonlinear effect in the untwisted disc (dashed line) is capable of regularising this behaviour and results in the disc circularising.}
\label{inviscid janosz comp}
\end{figure}

In order to determine the behaviour of $e$ we consider the strongly nonlinear limit ($q \rightarrow 1$) of our tightly wound theory,

\begin{equation}
F^{iso} (q) = \ln 2 - \sqrt{2} \sqrt{1 - q} + O(1 - q)
\label{iso q 1}
\end{equation}
The Functionals $I$ and J are given by,

\begin{equation}
I(q) = \frac{\pi}{2^{3/4}} (1 - q)^{-1/4} + O(1) , 
\end{equation}

\begin{equation}
J(q) = \frac{\pi}{2^{9/4}} (1 - q)^{-3/4} + O(1) , 
\end{equation}
we thus end up with a relationship between $I$ and $J$ in the nonlinear limit,

\begin{equation}
I \propto J^{1/3} .
\end{equation}
We can use this to obtain an expression for $q$ and $e$, as a function of $a$, near the marginally stable orbit,

\begin{equation}
1 - q \propto a^2 (a - r_s)^2 T(a)^{4/3} ,
\end{equation}

\begin{equation}
e \propto a^{-1} (a - r_s) T(a)^{-1/3} .
\end{equation}
So we see that as we approach the marginally stable orbit we approach an orbital intersection while $e$ diverges. This is a markedly different behaviour to that found in the simulations of \citet{Dewberry19b,Dewberry20}. So the explanation given in \citet{Dewberry20} for the reduction in eccentricity due to the nonlinear effects predicted in Paper I does not work, at least when applied to tightly wound waves. So what causes the effect seen in \citet{Dewberry19b,Dewberry20}? First, the tight-winding approximation assume that there is no reflection off the inner disc truncation. While the waves in \citet{Dewberry19b,Dewberry20} are twisted, it could be that there is a partial reflection off the disc truncation, which would likely strengthen the nonlinear effects. This means the explanation put forward in \citet{Dewberry20} may work for a partially reflected wave. Figure \ref{inviscid janosz comp} compares the envelopes of $q$ and $e$ for tightly-wound and untwisted waves highlighting the fundamentally different behaviour of $e$. These likely bracket the behaviour of partially reflected waves.

A second possibility which we shall explore in the next section is the wave is catastrophically damped near the marginally stable orbit in the presence of shear viscosity.

\subsection{Effect of including an $\alpha$-viscosity} \label{alpha visc janosz}

In Section \ref{ideal comp with dewberry} we obtained solutions for steady tightly-wound waves in an isothermal pseudo-Newtonian disc, which were compared against the findings of \citet{Dewberry19b} and \citet{Dewberry20}. To illustrate an application for the theory developed in this paper we now extend this calculation to include viscous damping, and compare the very different behaviours of shear and bulk viscosities on tightly wound waves. As we are considering an isothermal disc there is no phase shift in the 3D theory. 

The pseudo-Lagrangian for a steady-tightly wound wave in an isothermal disc in a Paczy\'{n}ski-Wiita potential, subject to a bulk $\alpha$-viscosity is

\begin{equation}
\mathcal{F}_a =  H^{\circ}_a \left( \frac{1}{2} \frac{q^2}{k^2} - F (q)\right) + \alpha_b a T_a^{\circ} (\varphi - \breve{\varphi}) F_{\rm bulk} (\breve{q}) \quad ,
\label{pseudo lagrangian bulk eg}
\end{equation}
while the same setup subject to a shear $\alpha-$viscosity is,

\begin{equation}
\mathcal{F}_a =  H^{\circ}_a \left( \frac{1}{2} \frac{q^2}{k^2} - F (q)\right) + \frac{2}{3} \alpha a T_a^{\circ} (\varphi - \breve{\varphi}) (F_{\rm shear} (\breve{q}) - F_{\rm bulk} (\breve{q})) .
\label{pseudo lagrangian shear eg}
\end{equation}
For an $\alpha$-disc $a T_a^{\circ} = H_a^{\circ}$. We shall restrict our attention to discs with a constant surface density profile, with truncation, so that $H_a^{\circ} = 2 \pi a c_0^2 \Sigma^{\circ} T(a)$, with $c_0$, $\Sigma^{\circ}$ constant sound speed and reference surface density respectively. As obtained in Section \ref{ideal comp with dewberry}, $N^2$ in this disc is

\begin{equation}
N^2 = \frac{2 n^2 a^4 r_g}{c_0^2 (a - 2 r_g)^3} \quad .
\end{equation}

The dispersion relation of the wave is independent of the viscosity and can be obtained by varying $F$ with respect to $q$. For an inward travelling wave (i.e. one propagating in from the outer boundary as in \citet{Dewberry19b} and \citet{Dewberry20}) this is given by

\begin{equation}
k = - \sqrt{\frac{q}{F^{\prime} (q)}} \quad .
\end{equation}
Varying Equations \ref{pseudo lagrangian bulk eg} and \ref{pseudo lagrangian shear eg} with respect to $\varphi$ and making use of the dispersion relation, we arrive at an equation describing how the nonlinearity varies in the disc. For bulk viscosity this is

\begin{equation}
a \partial_a \left( \frac{a H_a^{\circ}}{N} q^{1/2} F^{\prime} (q) \right) = \alpha_b a H_{a}^{\circ} F_{\rm bulk} (q) ,
\label{bulk eom}
\end{equation}
while for shear viscosity it is

\begin{equation}
a \partial_a \left( \frac{a H_a^{\circ}}{N} q^{1/2} F^{\prime} (q) \right) = \frac{2}{3} \alpha a H_{a}^{\circ} ( F_{\rm shear} (q)  - F_{\rm bulk} (q) ) \quad .
\label{shear eom}
\end{equation}
For an isothermal $\alpha$-disc we can evaluate the ``viscous integrals'',

\begin{equation}
F_{\rm bulk} (q) = 1 + \frac{1}{\sqrt{1 - q^2}} ,
\end{equation}

\begin{equation}
F_{\rm shear} (q) = \frac{3}{2} \frac{2 (1 - q^2)^{3/2} - 2 + 3 q^3}{(1 - q^2)^{3/2}} .
\end{equation}

\begin{figure}
\centering
\includegraphics[trim=0 0 0 0,clip,width=\linewidth]{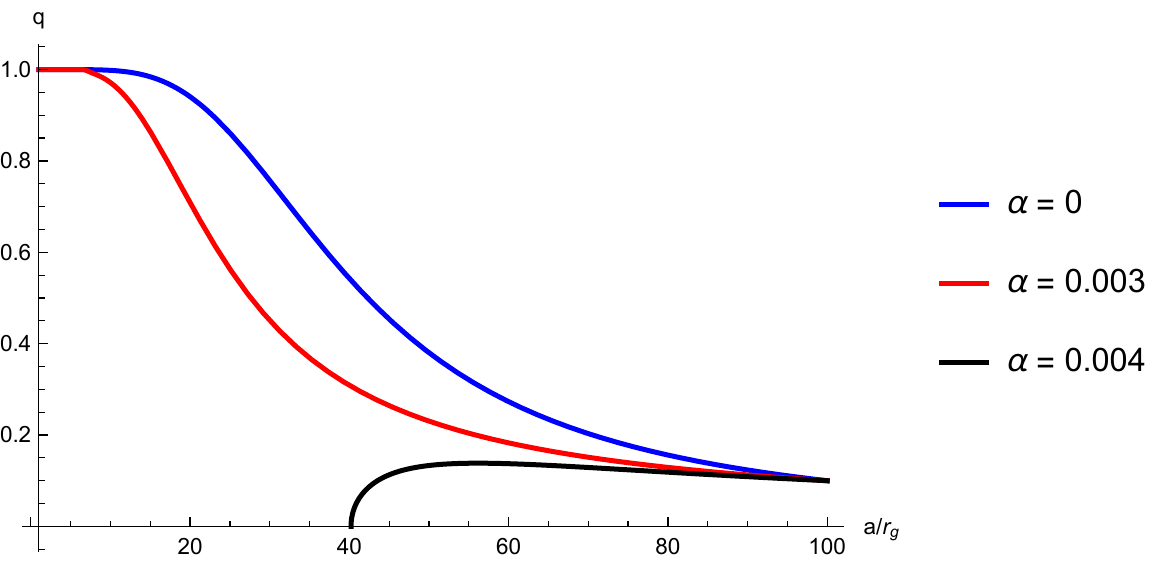}
\includegraphics[trim=0 0 0 0,clip,width=\linewidth]{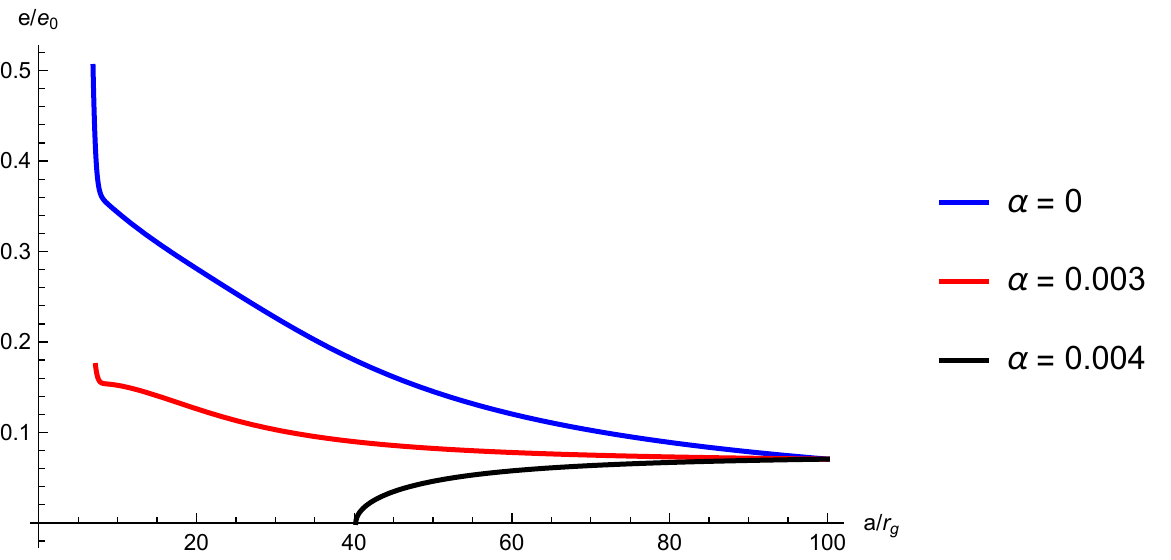}
\caption{Nonlinearity (top) and Eccentricity (bottom) for a tightly-wound wave approaching the marginally stable orbit in a truncated isothermal disc with bulk viscosity. As found by \citet{Ferreira09} when $\alpha_b$ is large enough the eccentric wave is entirely damped before it reaches the marginally stable orbit. When $\alpha_b$ is small enough that the wave makes it to the marginally stable orbit the bulk-viscosity is incapable of preventing the singular behaviour associated with the disc truncation.}
\label{bulk janosz comp}
\end{figure}

Equations \ref{bulk eom} and \ref{shear eom} can be solved using a numerical ODE solver. We solve these using Mathematica for $c_0 = 0.01$ and different values of $\alpha$/$\alpha_b$, with solutions shown in Figures \ref{bulk janosz comp} and \ref{shear janosz comp}. As expected the inclusion of bulk viscosity acts to damp the waves, with the nonideal solutions having consistently lower amplitudes than their ideal counterpart. For sufficiently large bulk viscosity the wave is completely damped before it reaches the marginally stable orbit, a result also found by \citet{Ferreira09} in linear theory. The solutions including shear viscosity excite the tightly wound wave throughout most of the disc, with the solutions containing a shear viscosity attaining large amplitudes at larger semimajor axis than their ideal counterpart. As the eccentric wave approaches the marginally stable orbit it becomes nonlinear enough for shear viscosity to instead damp the wave, resulting in a decline in the wave amplitude in the inner region of the disc.

For eccentric waves which make it to the marginally stable orbit, shear and bulk viscosities have very different effects on the wave, despite both damping the wave as $q$ approaches $1$. As we show in Appendix \ref{viscousity near isco}, bulk viscosity is incapable of regularising the singular behaviour associated with the disc truncation (see also Figure \ref{bulk janosz comp}). However the presence of a shear alpha-viscosity will regularise the singular behaviour associated with the truncation, even for arbitrarily small values of $\alpha$ (see Figure \ref{shear janosz comp} and Appendix \ref{viscousity near isco}). The fact that the tightly wound wave is catastrophically damped by shear viscosity, even when $\alpha$ is very small, potentially prevents any partial/total reflection. It is important to note that the wave remains strongly nonlinear as it approaches the disc truncation, so that the wave becomes increasingly tightly wound on the boundary. 

Many of the simulations carried out by \citet{Dewberry19b,Dewberry20} do not feature a decline in eccentricity as the wave approaches the marginally stable orbit, despite our calculations suggesting the wave should be damped to zero near this location. For very weak shear viscosity this damping is very abrupt and the eccentricity continues to increase inwards until very close to where the surface density falls to zero. Strictly, the very abrupt decline in the eccentricity seen when $\alpha$ is very small falls outside the scope of the short-wavelength theory as the lengthscale associated with viscous dissipation is comparable to the wavelength of the wave. The effects seen in the simulations of \citet{Dewberry19b,Dewberry20} were found even for 2D hydrodynamic simulations, where the fluid flow is laminar. This suggests the effect seen in these simulations may be dissipation of eccentricity due to the numerical viscosity of the grid when the waves become sufficiently nonlinear. In fact the inviscid theory suggests that the wave will never be resolved near the marginally stable orbit in an inviscid simulation as the nonlinearity will always increase inwards sufficiently to ensure the pressure gradients become limited by the grid resolution.

Finally Figure \ref{den etc janosz comp} shows plots of the surface density and radial velocity predicted by our theory, with $\alpha = 0.1$ and $c_0 = 0.01$. These appear qualitatively similar to the simulation outputs of \citet{Dewberry19b} and \citet{Dewberry20} and demonstrate the decline in eccentricity approaching the marginally stable orbit found in those simulations.

\begin{figure}
\centering
\includegraphics[trim=0 0 0 0,clip,width=\linewidth]{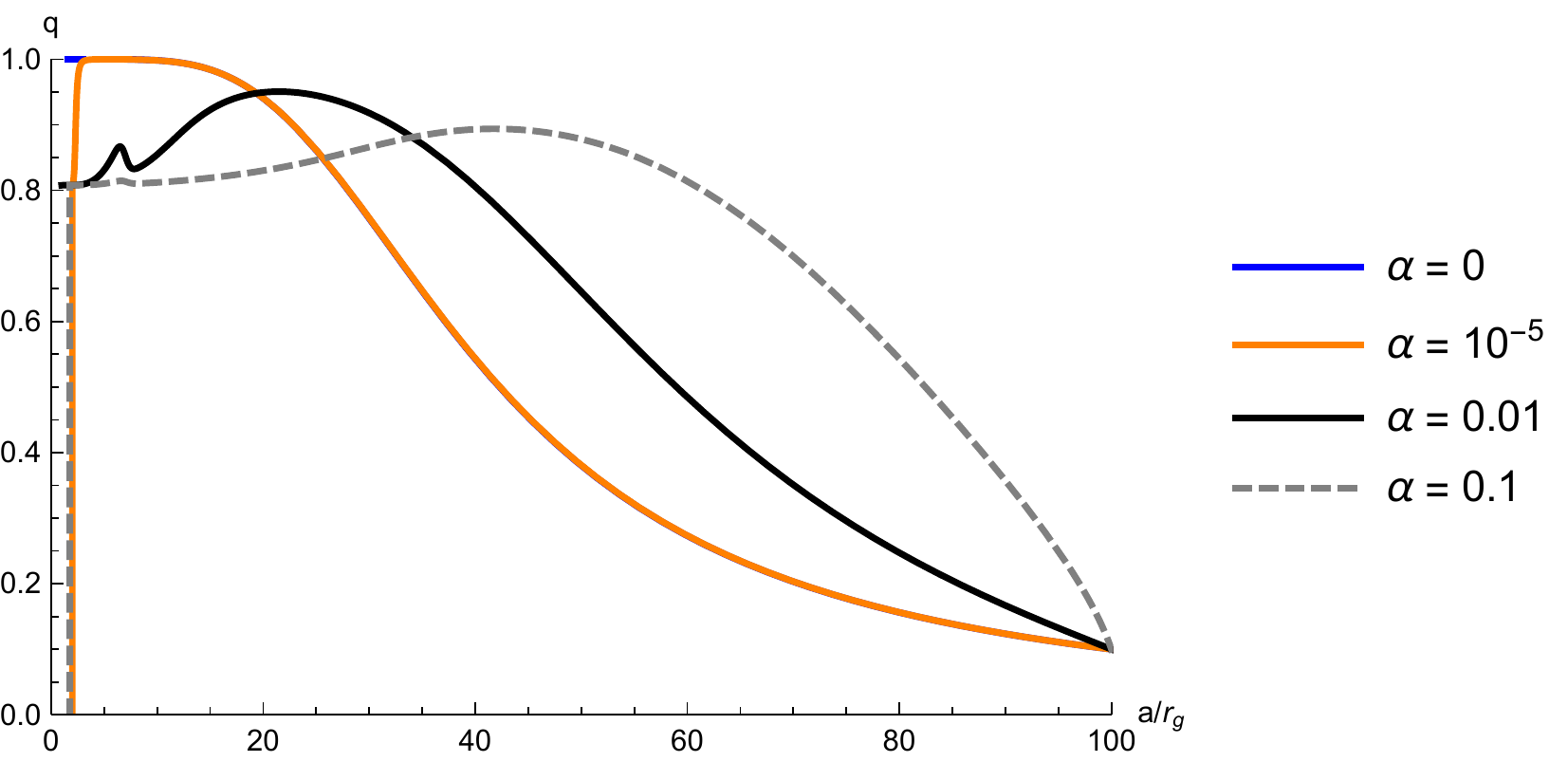}
\includegraphics[trim=0 0 0 0,clip,width=\linewidth]{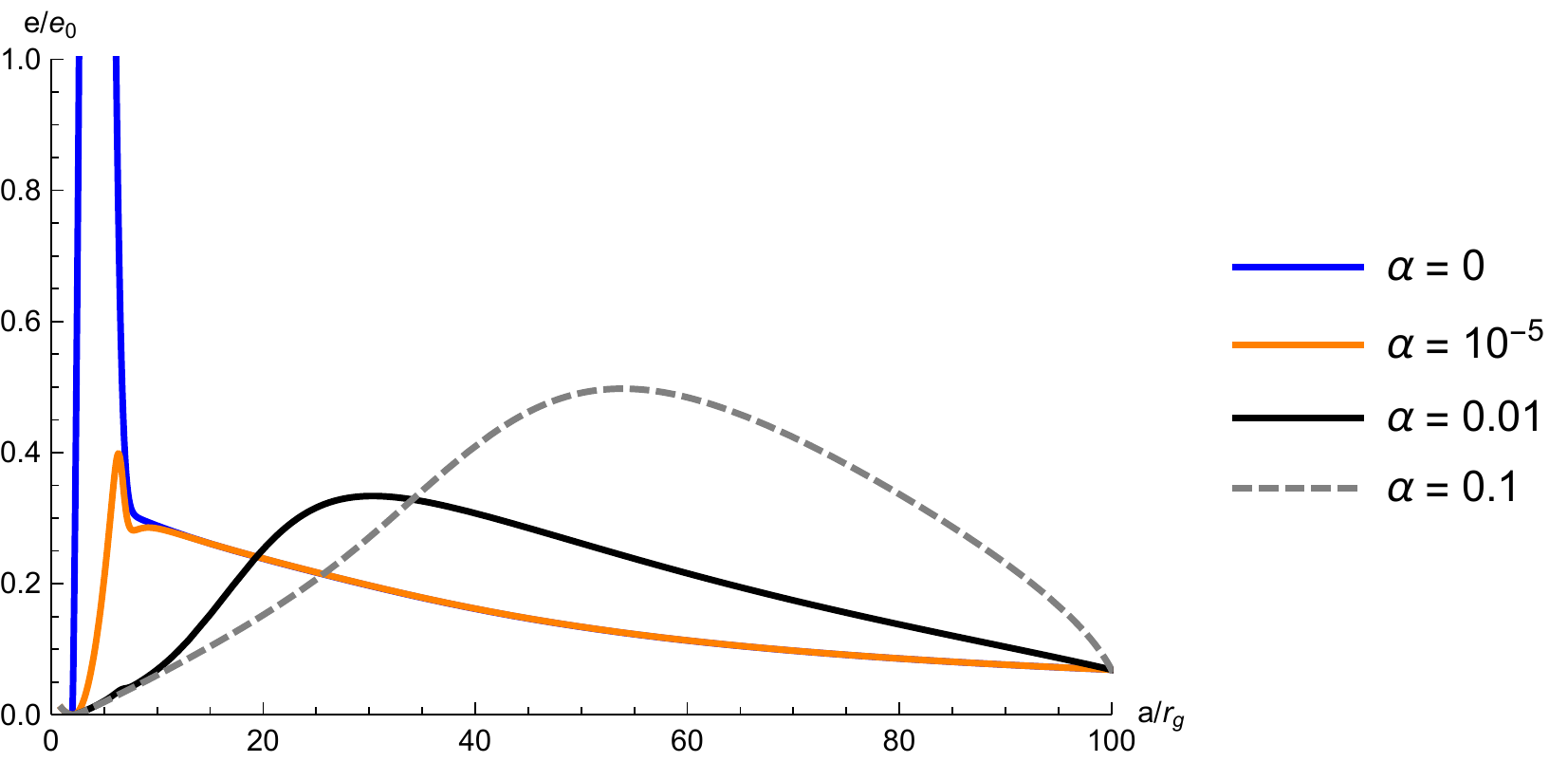}
\caption{Nonlinearity (top) and Eccentricity (bottom) for a tightly-wound wave approaching the marginally stable orbit in a truncated isothermal disc with shear viscosity. Unlike the bulk viscosity, the shear viscosity doesn't prevent the wave reaching the marginally stable orbit, as it is excites the wave in the linear limit. The shear viscosity becomes strongly dissipative as $q \rightarrow 1$ and this regularises the singular behaviour near the disc truncation.}
\label{shear janosz comp}
\end{figure}

\begin{figure*}
\begin{subfigure}{0.4\textwidth}
\includegraphics[trim=0 0 0 0,clip,width=0.75\linewidth]{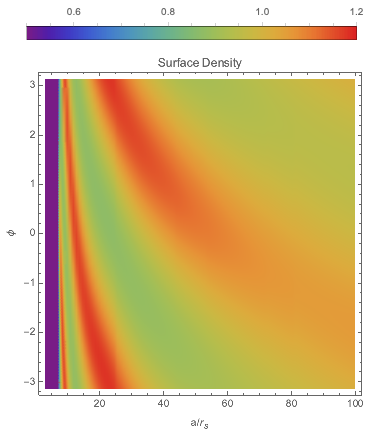}
\end{subfigure}
\begin{subfigure}{0.4\textwidth}
\includegraphics[trim=0 0 0 0,clip,width=0.75\linewidth]{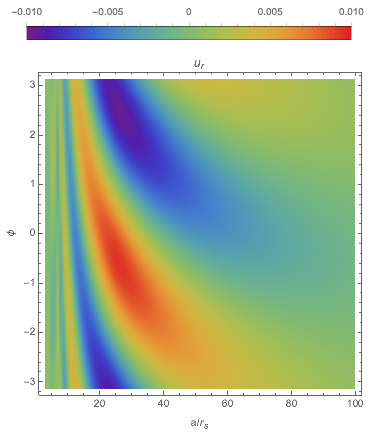}
\end{subfigure}

\begin{subfigure}{\textwidth}
\centering
\includegraphics[trim=0 0 0 0,clip,width=0.4\linewidth]{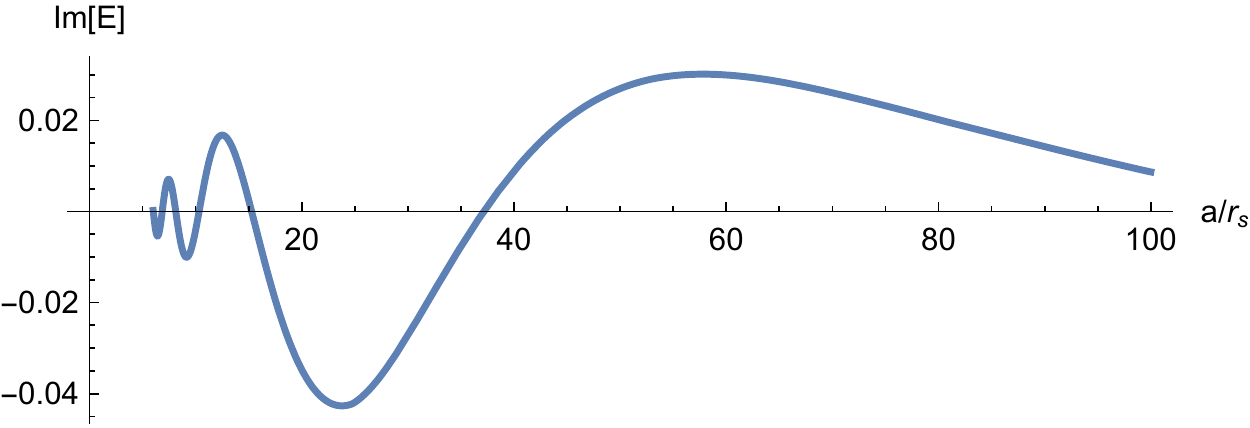}
\end{subfigure}
\setcounter{figure}{6}
\caption{Surface density (top left), radial velocity (top right) and imaginary part of the complex eccentricity (bottom) for a tightly-wound wave approaching the marginally stable orbit in a truncated isothermal disc with shear viscosity. The eccentric wave has been normalised so that the maximum eccentricity in the disc matches that in Figure A3. from \citet{Dewberry20}. }
\label{den etc janosz comp}
\end{figure*}

\section{Conclusions} \label{conc}

In this paper, I have formulated a nonlinear theory for highly twisted eccentric waves in astrophysical discs based on the averaged Lagrangian method of \citet{Whitham65} and its generalisation to include non-conservative terms in \citet{Jimenez76}. I obtain a dispersion relation which matches ideal theory, as the non-conservative terms are weak and do not appear at lowest order. Instead of the conserved fluxes found in Paper I one instead obtain an equation for the evolution of the angular momentum deficit where viscosity acts as a source/sink. I confirm that tightly-wound waves can be amplified by the viscous overstability and are excited by shear viscosity when they are of low enough amplitude. Shear viscosity damps sufficiently nonlinear tightly-wound waves, a result previously found in the planetary ring literature. Finally I discus the implications of this theory for the simulations of \citet{Dewberry20} and show that the decline of the eccentricity near the marginally stable orbit seen in their simulations cannot be caused by nonlinear steepening for a purely ingoing wave. However, it can be explained by either a partially reflected wave or catastrophic damping of the wave by shear viscosity in the strongly nonlinear limit.

\section*{Acknowledgements}

I am greatly indebted to Gordon Ogilvie under whose supervision much of the work presented in this paper was undertaken over the course of my PhD. Many of the ideas presented here originated in discussions we had during this time. I also wish to thank Janosz Dewberry for useful discussions, particularly in regards to the comparison of the theory presented here with recent simulations, and Guillaume Laibe for advise on revisions to this manuscript. This paper has been improved by several helpful suggestions by the anonymous reviewer, notably the suggestion to look at how damping of the eccentric wave affects the background accretion flow. 

I would like to thank the Science and Technologies Facilities Council (STFC) for funding this work through a STFC studentship, and the European Research Council (ERC). This research was supported by STFC through the grant ST/P000673/1 and the ERC through the CoG project PODCAST No 864965. This project has received funding from the European Union’s Horizon 2020 research and innovation program under the Marie Skłodowska-Curie grant agreement No 823823.

\section{Data availability}

The data underlying this article will be shared on reasonable request to the corresponding author.



\bibliographystyle{mnras}
\bibliography{tight_wound_eccentric} 




\appendix



\begin{strip}

\section{Disc self-gravity in the tight-winding limit.} \label{self gravity deriv}

\subsection{Derivation of the self-gravity potential in the tight-winding limit}

The disc self-gravity is described by Poisson's equation,

\begin{equation}
\nabla^2 \Phi_{\rm sg} = 4 \pi G \rho .
\end{equation}
We propose an asymptotic solution to Poisson's equation, valid within the disc, of the form

\begin{equation}
\Phi_{i} (R,\chi,\tilde{z}) = \Phi_{i 0}  (R,\chi) + H^2  \Phi_{i 2}  (R,\chi,\tilde{z}) + O(H^4) ,
\end{equation}
where $\chi$ is a local radial coordinate and $\tilde{z} = z/H$ is the stretched vertical coordinate. We assume $\Phi_{i 0}  (R,\chi) \approx \Phi_{l} (R)$ the contribution to the potential from matter a large distances. On the disc lengthscale $R$ the disc is approximately axisymmetric and we can calculate $\Phi_{l}$ at $z=0$ based on the mass distribution in the reference circular disc. Making use of the Green's function this means

\begin{equation}
\Phi_{l} (R,0) = - G \int_{r_{\rm in}}^{R}  \frac{1}{2 \pi} \int \left( x^2 + R^2 - 2 x R \cos \theta \right)^{-1/2} d \theta M_a(x) \, d x .
\end{equation}
For what follow we shall only need $\Phi_{l}$ to be defined on the disc surface ($z=0$). In order for the internal potential to satisfy Poisson's equation, we require

\begin{equation}
\left( \frac{1}{H^2} \frac{\partial^2}{\partial^2 \tilde{z}} +\frac{\partial^2}{\partial \chi^2} + \frac{1}{R} \frac{\partial}{\partial R} R \frac{\partial}{\partial R} \right) \Phi_{i} = 4 \pi G \rho (\chi, R) \tilde{\rho} (\tilde{z}) ,
\end{equation}
where we are treating the long and short radial coordinates ($R$ and $\chi$) as independent variables. We approximate the tightly-wound waves as being locally-axisymmetric. Substituting in the series expansion, multiplying by $H^2$ and Fourier transforming in $\chi$ we obtain

\begin{equation}
\frac{\partial^2 \hat{\Phi}_{i 2}}{\partial \tilde{z}^2} - (k H)^2 \hat{\Phi}_{i 2} + \frac{1}{R} \frac{\partial}{\partial R} \left( R \frac{\partial \Phi_{l}}{\partial R} \right) = 4 \pi G \hat{\rho} \tilde{\rho} ,
\end{equation}
which has a series solution, that, at leading order in $k H$, is

\begin{equation}
\Phi_{\rm d} = \Phi_e - \frac{1}{2 R} \frac{\partial}{\partial R} \left( R \frac{\partial \Phi_{l}}{\partial R} \right) z^2 + 4 \pi G H \Sigma \iint \tilde{\rho} \, d^2 \tilde{z} .
\end{equation}
The integration constants in the last term is zero, so that the local contribution to the disc potential vanishes when the density is zero.

\subsection{Derivation of the self-gravity contribution to the Lagrangian in the tight-winding limit} \label{self gravity lagrangian derivation}

The contribution to the Lagrangian from self gravity is

\begin{align}
\begin{split}
L_{\rm S G} &= - \int \Phi_{d} \, d m \\
&= - \int \frac{M_a}{2 \pi} \tilde{\rho} \Phi_{d} \, d a \, d \lambda \, d \tilde{z} \\
&= -\int M_a  \left\langle \Phi_{l} (R) \right\rangle \, d a + \int M_a \left\langle \frac{1}{2 R} \frac{\partial}{\partial R} \left( R \frac{\partial \Phi_{l}}{\partial R} \right) H^2  \right\rangle \, d a  -  2 G \lambda_{\rm sg} \int M_a^2 \langle H  J^{-1} \rangle \, d a   ,
\end{split}
\end{align}
where we define $\lambda_{sg} = \int_{-z_s}^{z_s} \tilde{\rho} \iint \tilde{\rho} d^2 \tilde{z} d \tilde{z}$. In the short wavelength limit we can approximate $\frac{1}{R} \frac{\partial}{\partial R} \left( R \frac{\partial \Phi_{l} (R)}{\partial R} \right) \approx \frac{1}{a} \frac{\partial}{\partial a} \left( a \frac{\partial \Phi_{l} (a)}{\partial a} \right)$ and expand the $\Phi_{l}$ term to second order in $e$ to obtain

\begin{equation}
L_{\rm S G} \approx \int \frac{1}{2} M_a  n a^2 \omega_{\rm sg} e^2 \, d a + \int \frac{1}{2} M_a \frac{1}{a} \frac{\partial}{\partial a} \left( a \frac{\partial \Phi_{l} (a)}{\partial a} \right) \langle H^2 \rangle \, d a  -  2 G \lambda_{\rm SG} \int M_a^2 \langle H  J^{-1} \rangle  \, d a    ,
\label{L SG in limit}
\end{equation}
where the forced precession due to self gravity is

\begin{equation}
\omega_{\rm sg} = - \frac{1}{2 n a^2} \frac{d}{d a} \left ( a^2 \frac{d \Phi_{l}}{d a} \right) .
\end{equation}
The affine disc Lagrangian with self gravity takes the form

\begin{equation}
L = \int \Sigma_0 \left[ \frac{1}{2} \left( u^2 + \dot{H}^2\right) - \Phi(\bar{x}) - \frac{1}{2} \Psi(\bar{x}) H^2 - \bar{\epsilon} \right] d^2 \bar{x}_0 + L_{\rm SG} .
\label{affine disc lagrangian}
\end{equation}
Varying this Lagrangian with respect to $H$ we obtain

\begin{equation}
\frac{\delta L}{\delta H} = \Sigma_0 \left[ - \ddot{H} - \Psi(\bar{x}) H + \frac{\hat{p}}{\Sigma} \right] + \frac{\delta L_{\rm SG}}{\delta H} .
\end{equation}
Setting this equal to zero, and substituting in Equation \ref{L SG in limit} for $L_{\rm SG}$, we obtain the evolutionary equation for $H$,
\begin{equation}
\ddot{H} = - \Psi(\bar{x}) H + \frac{\hat{p}}{\Sigma} + \frac{1}{a} \frac{\partial}{\partial a}\left( a \frac{\partial \Phi_{l}}{\partial a} \right) H - 4 \pi G \Sigma \lambda_{\rm sg} .
\label{dimensionfull scale height equation}
\end{equation}
Multiplying by $H$ and orbit averaging we obtain the virial relation,

\begin{equation}
\left \langle \dot{H}^2  - \Psi(\bar{x}) H^2 + \frac{P}{\Sigma} +\frac{1}{a} \frac{\partial}{\partial a} \left( a \frac{\partial \Phi_e}{\partial a} \right) H^2 - 4 \pi G \Sigma \lambda_{\rm sg} H \right \rangle = 0 .
\label{virial relation}
\end{equation}
Equation \ref{dimensionfull scale height equation} can be nondimensionalised to obtain the evolutionary equation for the dimensionless scale height $h$,

\begin{equation}
\frac{d^2 h}{d t^2} + \frac{n^2 h}{(1 - e \cos E)^3} = \frac{n^2}{j^{\gamma - 1} h^{\gamma}} - \psi_{\rm sg} h ,
\end{equation}
where we have introduced,

\begin{equation}
\psi_{\rm sg} = -\frac{1}{a} \frac{\partial}{\partial a} \left( a \frac{\partial \Phi_e}{\partial a} \right) + \frac{4 \lambda_{\rm sg}}{\bar{Q}^{\circ}} n^2 (j h)^{-1}  .
\end{equation}
Although we omit it here, a more careful treatment of the disc vertical structure in a self-gravitating disc shows that the disc self-gravity will excite inertial waves when the disc density varies with height in the disc. The inertial waves excited through self-gravity will combine with those expected to be present due to wave-coupling \citep{Kato08,Ferreira08,Dewberry18,Dewberry19}, likely leading to turbulence in the disc.

Substituting the virial relation (Equation \ref{virial relation}) into Equation \ref{affine disc lagrangian} and orbit averaging we obtain the secular Lagrangian for a tightly-wound wave in a fluid disc with self gravity,

\begin{equation}
L = L_{T} + \int H_{a}^{\circ} \left( \frac{1}{2} N^2 e^2 - F(q) - \frac{2 \lambda_{\rm s g}}{\bar{Q}^{\circ}} \langle j^{-1} h \rangle \right) \, d a ,
\end{equation}
where we have introduced a mass weighted vertically integrated Toomre-Q of the reference disc $\bar{Q} := \frac{1}{\Sigma^{\circ}} \int \rho^{\circ} Q^{\circ} d \tilde{z} = \frac{H^{\circ} n^2}{\pi G \Sigma^{\circ}}$. We have also introduced $N$, which for a disc with self-gravity, is given by

\begin{equation}
N^2 = \frac{M_a n a^2 ( \omega_{f} (a,0) + \omega_{\rm sg} (a) - \omega)}{H^{\circ}_{a}}  .
\end{equation}
Using the rescaled variables the Lagrangian becomes

\begin{equation}
L = L_{T} + \int H_{a}^{\circ} \left( \frac{1}{2} \frac{\tilde{e}^2}{k^2} - F(q) - \frac{2 \lambda_{\rm s g}}{\bar{Q}^{\circ}} \langle j^{-1} h \rangle \right) \, d a .
\end{equation}

\section{Derivation of the pseudo-Lagrangian for viscous stress model} \label{viscous damping and phase}

\subsection{Pseudo-Lagrangian from horizontal viscous stresses} \label{shear stress deriv}

In this Appendix we shall derive the contribution to the pseudo-Lagrangian from the viscosity in the absence of a phase shift. One could derive this pseudo-Lagrangian starting from the pseudo-Lagrangian for a viscous fluid. This will, however, necessitate accounting for the evolution of the orbit semimajor axis due to mass accretion as this contributes to the evolution of the eccentricity. Instead we opt to find a pseudo-Lagrangian that is consistent with \citet{Ogilvie14} and then taking the short wavelength limit.

From \citet{Ogilvie14} the eccentricity evolution caused by gradients in the stress tensor is given by

\begin{equation}
l \mathcal{M} \dot{\mathcal{E}}|_{\rm stress} = \int \int \left[2 (e^{i \phi} + \mathcal{E}) \partial_{\lambda} \left( J R^2 T^{\lambda \phi}\right) - i \lambda e^{i \phi} \partial_{\lambda} \left( J R_{\lambda} T^{\lambda \lambda} \right) - i e^{i \phi} J R^2 T^{\phi \phi} - \frac{J R^2}{\lambda} (e^{i \phi} + \mathcal{E} - \lambda \mathcal{E}_{\lambda}) T^{\lambda \phi} \right] d \phi d z ,
\label{shear visc contribution}
\end{equation}
in the $(\lambda,\phi)$ orbital coordinate system. In this coordinate system the short wavelength limit approximation is given by $e \ll 1$, $q \simeq 1$ where $q^2 \approx \lambda^2 \varphi_{\lambda}^2 \left(e_{\varphi}^2 + e^2 \varpi^2_{\varphi} \right)$. We have the following approximations:

\begin{align}
J &= \lambda j \approx \lambda (1 - q \cos (\phi - \phi_0)) + O(e) , \label{J approx} \\
R &\approx \lambda + O(e) , \label{R approx} \\
R_{\lambda} &\approx j + O(e) .
\end{align}

The viscous stress tensor is

\begin{equation}
T^{ij} = 2 \mu S^{ij} + \left( \mu_b - \frac{2}{3} \mu \right) \nabla \cdot u \, g^{i j}, 
\end{equation}
where $S^{ij}$ is the rate of strain tensor, $g^{i j}$ the inverse metric tensor and $\mu$ the dynamic viscosity. Define a vertically average dynamic shear viscosity $\bar{\mu} := \int \mu d z$ (similarly for bulk viscosity $\bar{\mu}_b := \int \mu_b d z$). Using the forms of $S^{ij}$ given in \citet{Ogilvie01} and keeping only the lowest order terms in the expansion we have vertically averaged stress tensors

\begin{align}
\bar{T}^{\lambda \lambda} &=  2 \bar{\mu} \frac{n}{j^2} \frac{ j_{\phi}}{j} + \left( \bar{\mu}_b - \frac{2}{3} \bar{\mu} \right) n \left( \frac{j_{\phi}}{j} + w \right) j^{-2} + O(e) , \label{stress lam lam} \\
\bar{T}^{\lambda \phi} &=  2 \bar{\mu} \frac{n}{\lambda} j^{-2} \left(\frac{1}{4} - j \right) + O(e) , \label{Tlamphi approx} \\
\bar{T}^{\phi \phi} &=  \left( \bar{\mu}_b - \frac{2}{3} \bar{\mu} \right) n \left( \frac{j_{\phi}}{j} + w \right) \lambda^{-2}+ O(e), \\
\bar{T}^{z z} &= 2 \bar{\mu} n w + \left( \bar{\mu}_b - \frac{2}{3} \bar{\mu} \right) n \left( \frac{j_{\phi}}{j} + w \right) + O(e) ,
\label{stress z z}
\end{align}
where we have introduced $w := \frac{h_{\phi}}{h}$.

Expanding out equation \ref{shear visc contribution} in terms of the short wavelength expansion, keeping only terms of lowest order, for shear viscosity and preforming the vertical average we get

\begin{equation}
l \mathcal{M} \dot{\mathcal{E}}|_{\rm stress} \approx \int \Biggl[2 e^{i \phi} \partial_{\lambda} \left(  \lambda^3 j \bar{T}^{\lambda \phi} \right) - i \lambda e^{i \phi} \partial_{\lambda} \left(\lambda j^2 \bar{T}^{\lambda \lambda} \right) - i e^{i \phi} j \lambda^3 \bar{T}^{\phi \phi}  - \lambda^2 j (e^{i \phi} - \tilde{\mathcal{E}}_{\varphi}) \bar{T}^{\lambda \phi} \Biggr] d \phi + O(e) .
\label{stress eq lambda}
\end{equation}
Rearranging to separate out the terms involving total derivatives we get

\begin{equation}
l \mathcal{M} \dot{\mathcal{E}}|_{\rm stress} \approx \partial_{\lambda} \int \lambda^2 j e^{i \phi}  \Biggl[2 \lambda \bar{T}^{\lambda \phi} - i j \bar{T}^{\lambda \lambda} \Biggr] d \phi + \int \lambda j  \Biggl[  i e^{i \phi} \left( j \bar{T}^{\lambda \lambda}  - \lambda^2 \bar{T}^{\phi \phi} \right)  - \lambda (e^{i \phi} - \tilde{\mathcal{E}}_{\varphi}) \bar{T}^{\lambda \phi} \Biggr] d \phi  \\
\end{equation}

Assuming that the vertically average dynamic viscosity is a function of pressure, surface density, scale height and semimajor axis only, then, in the short wavelength limit, it has the functional form $\bar{\mu} = \bar{\mu} (\lambda, q, \cos (\phi - \phi_{0}))$ (and the same for $\mu_b$) which we can use to simplify some of the orbit averages,

\begin{equation}
l \mathcal{M} \dot{\mathcal{E}}|_{\rm stress} \approx \partial_{\lambda}  \int \lambda^2 j e^{i \phi_0} \Biggl[2 \cos(\tilde{\phi}) \lambda \bar{T}^{\lambda \phi} + \sin(\tilde{\phi}) j \bar{T}^{\lambda \lambda} \Biggr] d \tilde{\phi} - e^{i \phi_0}  \int  \lambda j \Biggl[  \sin ( \tilde{\phi}) \left( j \bar{T}^{\lambda \lambda}  - \lambda^2 \bar{T}^{\phi \phi} \right)  + \lambda ( \cos( \tilde{\phi}) - q) \bar{T}^{\lambda \phi} \Biggr] d \tilde{\phi} ,
\end{equation}
where we have shifted the phase of the orbit average by $\phi_0$ and introduced $\tilde{\phi} := \phi - \phi_0$. We now switch to the $a,E$ variables used in the rest of the work. At this order, $\lambda \approx a$, $\tilde{\phi} \approx E$ (strictly shifted $E$), $\phi_0 \approx E_0 + \varpi$ and $l \mathcal{M} \approx M_a n a^2$ and (after multiplying both sides by $i/2$) equation \ref{stress eq lambda} becomes

\begin{align}
\begin{split}
\frac{i}{2} M_a n a^2 \dot{\mathcal{E}}|_{\rm stress} &\approx \frac{\partial}{\partial a} \frac{i}{2} a^2 e^{i (E_0+\varpi)} \int j \Biggl[2 \cos E \, a \bar{T}^{\lambda \phi} + \sin E \, j \bar{T}^{\lambda \lambda} \Biggr] d E \\
&- \frac{i}{2} e^{i (E_0+\varpi)}  \int  a j \Biggl[  \sin E  \, \left( j \bar{T}^{\lambda \lambda}  - a^2 \bar{T}^{\phi \phi} \right)  + a ( \cos E - q) \bar{T}^{\lambda \phi} \Biggr] d E .
\label{shear equation}
\end{split}
\end{align}
The left hand side comes from a variation of the Lagrangian, $L_T$, with

\begin{equation}
\frac{\delta L_T}{\delta \mathcal{E}^{*}} = \frac{i}{2} M_a n a^2 \dot{\mathcal{E}} \quad ,
\end{equation}
so we are searching for a pseudo-Lagrangian $F$ so that varying $-F$ with respect to $\mathcal{E}^{*}$ yields the right hand side of Equation \ref{shear equation}. Noting that

\begin{equation}
\frac{\partial}{\partial \mathcal{E}^{*}_{a}} \frac{\mathrm{Re}[i a \mathcal{E}^{*}_a \breve{\mathcal{E}}_a]}{\breve{q}}= \frac{i}{2} e^{i (\breve{E}_0+\breve{\varpi})} ,  \quad \frac{\partial}{\partial \mathcal{E}^{*}} \frac{\mathrm{Re}[i a \mathcal{E}^{*} \breve{\mathcal{E}}_a]}{\breve{q}} = \frac{i}{2} e^{i (\breve{E}_0 + \breve{\varpi})},
\end{equation}
we obtain a pseudo Lagrangian density,

\begin{equation}
F_a \approx a^2 \frac{\mathrm{Re}[i \mathcal{E}^{*}_a \breve{\mathcal{E}}_a]}{\breve{q}} \int \breve{j} \Biggl[2 \cos E \, a \breve{\bar{T}}^{\lambda \phi} + \sin E \, \breve{j} \breve{\bar{T}}^{\lambda \lambda} \Biggr] d E + \frac{\mathrm{Re}[i \mathcal{E}^{*} \breve{\mathcal{E}}_a]}{\breve{q}}  \int  a \breve{j} \Biggl[  \sin E  \, \left( \breve{j} \breve{\bar{T}}^{\lambda \lambda}  - a^2 \breve{\bar{T}}^{\phi \phi} \right)  + a ( \cos E - \breve{q}) \breve{\bar{T}}^{\lambda \phi} \Biggr] d E ,
\end{equation}
where the breves ($\breve{\cdot}$) denote dependence on the complimentary phase space variables. We now preform the expansion of the complimentary phase variable $\breve{\varphi}$ about $\varphi$. We make use of the following expansions:

\begin{equation}
\frac{\mathrm{Re} [i \mathcal{E}_a^{*} \breve{\mathcal{E}}_a]}{\breve{q}} = - \frac{q}{a} \sin(\breve{E}_0 + \breve{\varpi} - E_0 - \varpi) \approx \frac{\breve{q}}{a} (\varphi - \breve{\varphi}) ( \breve{\varpi}_{\breve{\varphi}} + \breve{E}_{0,\breve{\varphi}}) ,
\end{equation}

\begin{equation}
\frac{\mathrm{Re} [i \mathcal{E}^{*} \breve{\mathcal{E}}_a]}{\breve{q}} = -e \sin (\breve{E}_0 + \breve{\varpi} - \varpi) \approx -e \sin \breve{E}_0 - (\varphi - \breve{\varphi}) \breve{e}_{\breve{\varphi}} \sin \breve{E}_0 + (\varphi - \breve{\varphi}) \breve{e} (\breve{\varpi}_{\breve{\varphi}} + \breve{E}_{0,\breve{\varphi}}) \cos \breve{E}_0 ,
\end{equation}
from which we obtain an approximation for the (averaged) pseudo-Lagrangian density in the short wavelength limit,

\begin{align}
\begin{split}
\mathcal{F}_a &\approx a \breve{q} (\varphi - \breve{\varphi}) ( \breve{\varpi}_{\breve{\varphi}} + \breve{E}_{0,\breve{\varphi}}) \int \breve{j} \Biggl[2 \cos E \, a \breve{\bar{T}}^{\lambda \phi} + \sin E \, \breve{j} \breve{\bar{T}}^{\lambda \lambda} \Biggr] d E \\
& - (\varphi - \breve{\varphi})  \left[ \breve{e}_{\breve{\varphi}} \sin \breve{E}_0 - \breve{e} (\breve{\varpi}_{\breve{\varphi}} + \breve{E}_{0,\breve{\varphi}}) \cos \breve{E}_0 \right]  \int  a \breve{j} \Biggl[  \sin E  \, \left( \breve{j} \breve{\bar{T}}^{\lambda \lambda}  - a^2 \breve{\bar{T}}^{\phi \phi} \right)  + a ( \cos E - \breve{q}) \breve{\bar{T}}^{\lambda \phi} \Biggr] d E ,
\end{split}
\end{align}
where we have dropped the $O(1)$ term as it doesn't contribute to the dynamics. In the tight winding limit $\breve{e}_{\breve{\varphi}} = \breve{E}_{0,\breve{\varphi}} = 0$ and $E_0 = \pi/2$, so the terms which did not originate in the total derivative vanish. Substituting the expressions for the stress tensor (Equations \ref{stress lam lam}-\ref{stress z z}) the pseudo-Lagrangian density becomes

\begin{equation}
\mathcal{F}_a \approx 2 n a \breve{q} (\varphi - \breve{\varphi}) \breve{\varpi}_{0,\breve{\varphi}} \int \breve{\bar{\mu}} \breve{j}^{ -1} \Biggl[2 \cos E \,   \left(\frac{1}{4} - \breve{j} \right) + \sin E \,   \breve{j}_{E}  \Biggr] d E + n a \breve{q} (\varphi - \breve{\varphi}) \breve{\varpi}_{0,\breve{\varphi}} \int \left( \breve{\bar{\mu}}_b - \frac{2}{3} \breve{\bar{\mu}} \right) \left( \frac{\breve{j}_{E}}{\breve{j}} + \breve{w} \right)  \sin E \, d E .
\end{equation}

Proposing that the viscosity has the functional form $\bar{\mu}_b = \frac{\alpha_b}{2 \pi}  \mu^{\circ}_a (a) m(q,\cos E)$, $\bar{\mu} = \frac{\alpha}{2 \pi} \mu^{\circ}_a (a) m(q,\cos E)$, appropriate for a viscosity which is a power law of pressure, surface density and scale height. Introducing $T_a^{\circ} = n \mu_a^{\circ}$ then, integrating over the disc, we can write the pseudo-Lagrangian as

\begin{equation}
\mathcal{F} = \frac{2}{3} \alpha \int a T_a^{\circ} (\varphi - \breve{\varphi}) \breve{\varpi}_{0,\breve{\varphi}} F_{\rm shear} (\breve{q}) \, d a + \left( \alpha_b - \frac{2}{3} \alpha \right) \int a T_a^{\circ} (\varphi - \breve{\varphi}) \breve{\varpi}_{0,\breve{\varphi}} F_{\rm bulk} , (\breve{q}) \, d a
\end{equation}
where we have introduced

\begin{equation}
F_{\rm bulk} (q) := \frac{q}{2 \pi} \int_0^{2 \pi} \tau (q,\cos E) \sin^2 E \, d E ,
\end{equation}

\begin{equation}
F_{\rm shear} (q) := \frac{3 q}{2 \pi} \int_0^{2 \pi} m (q,\cos E) j^{-1} \Biggl[2 \cos E \,   \left(\frac{1}{4} - j \right) + q \sin^2 E \,  \Biggr] d E ,
\end{equation}
with

\begin{equation}
\tau (q,\cos E) := m(q,\cos E) \left( \frac{q}{j} - \frac{1}{h} \frac{\partial h}{\partial \cos E} \right) .
\end{equation}
For 2D or isothermal discs these integrals can be evaluated in terms of Legendre functions, similar to the 2D eccentric disc Hamiltonian of \citet{Ogilvie19}. In the $\tilde{e} = q$ gauge the pseudo-Lagrangian is

\begin{equation}
\mathcal{F} = \frac{2}{3} \alpha \int a T_a^{\circ} (\varphi - \breve{\varphi}) F_{\rm shear} (\breve{q}) \, d a + \left( \alpha_b - \frac{2}{3} \alpha \right) \int a T_a^{\circ} (\varphi - \breve{\varphi}) F_{\rm bulk} (\breve{q})  \, d a \quad .
\end{equation}
Interestingly when $F_{\rm bulk} (q) > F_{\rm shear} (q)$ the shear viscosity excites the eccentric wave and we have a form of viscous overstability.

\subsection{Pseudo-Lagrangian for the phase shift.} \label{phase shift section}

Viscous terms in a fluid theory can be included by making use of the following Pseudo-Lagrangian,

\begin{equation}
F_{D} = \iiint \breve{J}_3 x_i \breve{\nabla}_{j} \breve{T}^{i j} d^3 x_0 .
\end{equation}
Splitting the stress tensor into a horizontal and vertical part,

\begin{equation}
T^{i j} = T_{h}^{i j} + T^{z z} \delta_z^i \delta_z^j ,
\end{equation}
This leads to the following pseudo-Lagrangian density from viscous terms

\begin{equation}
 F_{D , \, a} = \int \breve{J}_3 x_i \nabla_{j} \breve{\tau}_h^{i j} d \phi + \iint \breve{J}_3 z \frac{\partial \breve{T}^{z z}}{\partial \breve{z}} d \phi d z ,
\end{equation}
where we are working with the $(\lambda, \phi)$ coordinate system as it is more convenient when dealing with viscous terms. The contribution from the horizontal viscous stresses is obtained from

\begin{equation}
 F_{\rm V, \, a} := \int \breve{J}_3 x_i \breve{\nabla}_{j} \breve{\bar{T}}_h^{i j} d \phi ,
\end{equation}
an expression for which is obtained in Appendix \ref{shear stress deriv}. The contribution from the phase shift is obtain from

\begin{equation}
 F_{\rm P, \, a} := \iint \breve{J}_3 z \frac{\partial \breve{T}^{z z}}{\partial \breve{z}} d \phi d z .
\label{F phase def}
\end{equation}
Expanding z about $\tilde{z}$,

\begin{equation}
 z \approx \breve{z} + \frac{H - \breve{H}}{\breve{H}} \breve{z} ,
\end{equation}
substituting into Equation \ref{F phase def} and integrating by parts we arrive at

\begin{equation}
 F_{\rm P, \, a} = - \int_{0}^{2 \pi} \frac{H - \breve{H}}{\breve{H}} \breve{J} \breve{\bar{T}}^{z z} d \lambda ,
\end{equation}
where we have dropped a term that only depends on the complementary phase space which has no influence on the dynamics.

In the short wavelength limit we have $H = H(q ,\sin(\phi - \varpi(\varphi)))$. Expanding $\varphi$ about $\breve{\varphi}$ we have

\begin{equation}
H \approx \breve{H} - \left(\varphi - \breve{\varphi} \right) \breve{\varpi}_{\breve{\varphi}} \breve{H}_{\phi} .
\end{equation}
Thus the averaged pseudo-Lagrangian contribution, $\mathcal{F}_{\rm P, a}$, is given by

\begin{equation}
 \mathcal{F}_{\rm P, \, a} \approx  \left(\varphi - \breve{\varphi} \right) \breve{\varpi}_{\breve{\varphi}} \lambda \int_{0}^{2 \pi} w \breve{j} \breve{\breve{T}}^{z z} d \phi ,
\end{equation}
where $w = h_{\phi}/h$. In short wavelength limit, $\bar{T}^{z z}$ is given by Equation \ref{stress z z}. Switching back to the $(a,E)$ coordinate system, we assume that the vertically integrated viscosity has the functional form $\bar{\mu}_b = \frac{\alpha_b}{2 \pi}  \mu^{\circ}_a (a) m(q,\cos \tilde{E})$, $\bar{\mu} = \frac{\alpha}{2 \pi} \mu^{\circ}_a (a) m(q,\cos E)$, and introduce $T^{\circ}_{a} = \mu^{\circ}_{a} n$. The phase shift term is then given by

\begin{equation}
 \mathcal{F}_{\rm P, \, a} =  \left(\varphi - \breve{\varphi} \right) \breve{\varpi}_{\breve{\varphi}} a T^{\circ}_{a}  \left[ 2 \alpha F_{\rm phase} (\breve{q}) - \left( \alpha_b - \frac{2}{3} \alpha \right) F_{\rm phase \, b} (\breve{q})  \right]  ,
\end{equation}
where we have introduced,

\begin{equation}
F_{\rm phase} (q) = \frac{1}{2 \pi} \int_{0}^{2 \pi} m w^2 j \, d \phi  = \frac{1}{2 \pi} \int_{0}^{2 \pi} m w^2 j \, d E ,
\end{equation}

\begin{equation}
F_{\rm phase \, b} (q) = -\frac{1}{2 \pi} \int_{0}^{2 \pi} m j \left( \frac{j_{\phi}}{j} + w \right) w \, d \phi = -\frac{1}{2 \pi} \int_{0}^{2 \pi} m j \left( \frac{j_{E}}{j} + w \right) w \, d E  .
\end{equation}
In the $\tilde{e} = q$ gauge of the tightwinding limit $\breve{\varpi}_{\breve{\varphi}} = 1$. We thus arrive at an expression for $\mathcal{F}_{D a}$ (in $a,E$ variables),

\begin{equation}
 \mathcal{F}_{D, \, a} = \left(\varphi - \breve{\varphi} \right) a T^{\circ}_{a}  \left\{ \frac{2}{3} \alpha \left[ F_{\rm shear} (\breve{q}) + 3 F_{\rm phase} (\breve{q}) \right] + \left( \alpha_b - \frac{2}{3} \alpha \right) \left[ F_{\rm bulk} (\breve{q}) -  F_{\rm phase \, b} (\breve{q}) \right] \right \} ,
\end{equation} 
where $F_{\rm shear}$ and $F_{\rm bulk}$ are given in Appendix \ref{shear stress deriv}.

\subsection{Effect of the eccentric wave on the accretion flow.} \label{eccentric accretion flow}

From \citet{Ogilvie14}, the viscous torque in an eccentric disc, in the $(\lambda,\phi)$ coordinate system, is

\begin{equation}
 \mathcal{G} = - \iint J R^{2} T^{\lambda \phi} \, d \phi \, d z .
\end{equation} 
Making use of the short-wavelength approximations introduced in Section \ref{shear stress deriv} (Equations \ref{J approx}, \ref{R approx} and \ref{Tlamphi approx}), the viscous torque in the short-wavelength limit is

\begin{equation}
 \mathcal{G} = - 2 \lambda^{2} n \int \bar{\mu} j^{-1} \left(\frac{1}{4} - j \right) d \phi .
\end{equation}
Assuming the functional form for viscosity, $\bar{\mu} = \frac{\alpha}{2 \pi} \mu_a^{\circ} (a) m (q, \cos E)$, and switching to the $(a,E)$ coordinate system we arrive at

\begin{equation}
 \mathcal{G} = - 2 \alpha a^2 T_a^{\circ} \frac{1}{2 \pi} \int m j^{-1} \left(\frac{1}{4} - j \right)\, d E .
\end{equation}
This viscous torque is responsible for driving the background accretion flow in the disc.

\section{Effect of viscosity near the disc truncation} \label{viscousity near isco}

In this appendix we consider the strongly nonlinear ($q \rightarrow 1$) limit of the viscous, isothermal disc models considered in Section \ref{alpha visc janosz}. In the strongly nonlinear limit the viscous integrals, for an isothermal disc, simplify to

\begin{equation}
F_{\rm bulk} (q) = 2^{-1/2} (1 - q)^{-1/2} + O(1) , 
\label{bulk q 1}
\end{equation}

\begin{equation}
F_{\rm shear} (q) = \frac{3}{2} 2^{-3/2} (1 - q)^{-3/2} + O(1) .
\label{shear q 1}
\end{equation}

Introducing a new variable $Y = \frac{a H_a^{\circ}}{N} (1 - q)^{-1/2}$, we can use Equations \ref{iso q 1}, \ref{bulk q 1} and \ref{shear q 1} to rewrite the equation describing how $q$ varies in the disc in the presence of bulk viscosity  (Equation \ref{bulk eom}) as

\begin{equation}
a \frac{\partial Y}{\partial a} \approx \alpha_b N Y,
\end{equation}
while the equation with shear viscosity (Equation \ref{shear eom}) becomes

\begin{equation}
a \frac{\partial Y}{\partial a} \approx \frac{1}{2} \alpha N^3 ( a H_a^{\circ})^{-2} Y^3 ,
\end{equation}
These can be directly integrate to obtain

\begin{equation}
Y = Y_0 \exp \left( \alpha_b \int \frac{N}{a} d a \right) ,
\end{equation}
for a disc with bulk viscosity and

\begin{equation}
Y^{-2} = C - \alpha \int \frac{N^3}{a} (a H_a^{\circ})^{-2} d a,
\end{equation}
for a disc with shear viscosity, with $Y_0$ and $C$ constants. Changing back to $q$ we obtain

\begin{equation}
1 - q = \frac{(a H_a^{\circ})^2}{N^2} Y_0^{-2} \exp \left( -2 \alpha_b \int \frac{N}{a} d a \right) ,
\end{equation}
for the bulk viscosity and

\begin{equation}
1 - q = \frac{(a H_a^{\circ})^2}{N^2} \left[ C - \alpha \int \frac{N^3}{a} (a H_a^{\circ})^{-2} d a \right] ,
\end{equation}
for the shear viscosity. Because $F_{\rm shear} (q)$ diverges faster than $F_{\rm bulk} (q)$ as $q \rightarrow 1$ the latter solution also applies to a disc with a mix of bulk and shear viscosity.  In both cases when $H_a \rightarrow 0$, $q \rightarrow 1$ as expected, however these do so at differing rates. Substituting this into the Expression for $I(q)$ in the strongly nonlinear limit we obtain expressions for $e$ in the strongly nonlinear limit. For a disc with only bulk viscosity

\begin{equation}
e \propto (a H_a^{\circ})^{-1/4} N^{-3/2} \exp \left( - \frac{\alpha_b}{4} \int \frac{N}{a} d a\right),
\end{equation}
while a disc with shear viscosity has
 
\begin{equation}
e \propto (a H_a^{\circ})^{-1/4} N^{-3/2} \left[ C - \alpha \int \frac{N^3}{a} (a H_a^{\circ})^{-2} d a \right]^{1/8},
\end{equation}

For a disc with only bulk viscosity there is an additional exponential suppression of $e$ due to the presence of bulk viscosity. This remains bounded as we approach the disc truncation so cannot regularise the behaviour as $a H_a^{\circ} \rightarrow 0$. For a disc with shear viscosity, as we approach the truncation, the term proportional to $\alpha$ will grow until it approaches $C$ so that $e \rightarrow 0$. For all values of $\alpha$, the $\alpha \int \frac{N^3}{a} (a H_a^{\circ})^{-2} d a $ term will reach $C$ before $a H_a^{\circ}$ reaches zero so the eccentricity of a tightly-wound wave approaching a disc truncation is always bounded and will vanish, even for arbitrarily small (but non-zero) shear viscosity. For very weak shear viscosity the disc truncation can still cause the eccentricity to grow to large (but finite) values before it is catastrophically damped by the viscosity. However as the eccentricity remains bounded, and it's absolute value is arbitrary in the short wavelength limit, it can always be rescaled so that we remain within the domain of validity of the short-wavelength theory.

\end{strip}



\bsp	
\label{lastpage}
\end{document}